\documentclass[aps,prapplied,reprint,superscriptaddress]{revtex4-2}
\usepackage{graphicx}
\usepackage{color,amssymb,amsmath}
\usepackage[normalem]{ulem}
\usepackage{hyperref}
\usepackage{lipsum}
\usepackage{physics}
\usepackage{float}

\makeatletter
\newcommand*{\rom}[1]{\expandafter\@slowromancap\romannumeral #1@}
\makeatother
\newcommand{\fr}{f_{\mathrm{r}}}
\newcommand{\fd}{f_{\mathrm{d}}}
\newcommand{\Qi}{Q_{\mathrm{i}}}
\newcommand{\Qe}{Q_{\mathrm{e}}}

\newcommand{\VG}{V_{\mathrm{G}}}

\begin{document}
\title{Flip-chip-based microwave spectroscopy of Andreev bound states in a planar Josephson junction}

\author{M.~Hinderling}
\affiliation{IBM Research Europe - Zurich, S\"aumerstrasse 4, 8803 R\"uschlikon, Switzerland}

\author{D.~Sabonis}
\affiliation{IBM Research Europe - Zurich, S\"aumerstrasse 4, 8803 R\"uschlikon, Switzerland}

\author{S.~Paredes}
\affiliation{IBM Research Europe - Zurich, S\"aumerstrasse 4, 8803 R\"uschlikon, Switzerland}

\author{D.~Z.~Haxell}
\affiliation{IBM Research Europe - Zurich, S\"aumerstrasse 4, 8803 R\"uschlikon, Switzerland}
	
\author{M.~Coraiola}
\affiliation{IBM Research Europe - Zurich, S\"aumerstrasse 4, 8803 R\"uschlikon, Switzerland}
	
\author{S.~C.~ten~Kate}
\affiliation{IBM Research Europe - Zurich, S\"aumerstrasse 4, 8803 R\"uschlikon, Switzerland}

\author{E. Cheah}
\affiliation{Solid State Physics Laboratory, ETH Zurich, Otto-Stern-Weg 1, 8093 Z\"urich, Switzerland}

\author{F. Krizek}
\affiliation{IBM Research Europe - Zurich, S\"aumerstrasse 4, 8803 R\"uschlikon, Switzerland}
\affiliation{Solid State Physics Laboratory, ETH Zurich, Otto-Stern-Weg 1, 8093 Z\"urich, Switzerland}

\author{R. Schott}
\affiliation{Solid State Physics Laboratory, ETH Zurich, Otto-Stern-Weg 1, 8093 Z\"urich, Switzerland}

\author{W.~Wegscheider}
\affiliation{Solid State Physics Laboratory, ETH Zurich, Otto-Stern-Weg 1, 8093 Z\"urich, Switzerland}

\author{F.~Nichele}
\email{fni@zurich.ibm.com}
\affiliation{IBM Research Europe - Zurich, S\"aumerstrasse 4, 8803 R\"uschlikon, Switzerland}

\date{\today}

\begin{abstract}
We demonstrate a flip-chip-based approach to microwave measurements of Andreev bound states in a gate-tunable planar Josephson junction using inductively-coupled superconducting low-loss resonators. By means of electrostatic gating, we present control of both the density and transmission of Andreev bound states. Phase biasing of the device shifted the resonator frequency, consistent with the modulation of supercurrent in the junction. Two-tone spectroscopy measurements revealed an isolated Andreev bound state consistent with an average induced superconducting gap of 184~$\mu$eV and a gate-tunable transmission approaching $0.98$. Our results represent the feasibility of using the flip-chip technique to address and study Andreev bound states in planar Josephson junctions, and they give a promising path towards microwave applications with superconductor-semiconductor two-dimensional materials. 
\end{abstract}

\maketitle

\section{\label{sec:level1}INTRODUCTION}
A class of gate-tunable superconducting-semiconducting materials have recently been investigated as potential building blocks for hybrid amplifiers~\cite{JavadAmplifier}, tunable quantum buses~\cite{CasparisQuantumBus}, non-reciprocal devices~\cite{BlaisNonreciprocal} and facilitation of integration with cryogenic control systems~\cite{PaukaCryogenicCMOS}. Moreover, hybrid superconducting devices based on Andreev bound states (ABSs)~\cite{Andreev1964} were realized as gate- and flux-tunable two-level system \cite{Bretheau2013, SupercurrentSpectroscopy, AtomicContact1, GatemonDelft, NanowireLongJunction1, GateTunableFluxonium, NanowireLongJunction3, NanowireLongJunction5} 
and qubits \cite{AtomicContact2, GatemonCopenhagen, HaysShortJunction, LuthiTransmon,HaysAndreevSpinQubit,  SabonisGatemon2022, SpinTransmonQubit}, potentially enabling long coherence times~\cite{ProtectedQubitCopenhagen,ProtectedQubitGynis}. Such ABS-based qubits~\cite{Desposito,AndreevLevelQubit,AndreevSpinQubit} have been realized in superconducting atomic contact break-junctions~\cite{AtomicContact2}, hybrid Josephson junctions (JJs) using nanowires~\cite{HaysShortJunction,HaysAndreevSpinQubit}, graphene~\cite{WangGatemonGraphene} and two-dimensional hybrid systems~\cite{CasparisGatemon2DEG}. 

\raggedbottom

Two-dimensional superconductor-semiconductor heterostructures grown on III-V substrates, such as InP~\cite{GateDependentIC}, offer several advantages in terms of reproducibility and scalability, but pose difficulties in realizing high-quality microwave components, e.g. coplanar waveguide (CPW) resonators. This is due to, for example, the large dielectric losses in the host structure~\cite{Resonator2DEG2,Resonator2DEG3,Resonator2DEG4,Resonator2DEG5} as well as to the piezoelectric properties of III-V materials~\cite{Piezoelectricity}. Furthermore, techniques used in the fabrication of microwave components degrade semiconductor quality. Attempts to overcome these limitations included removal of the buffer layers in the area surrounding the device~\cite{CasparisGatemon2DEG}, or by selective area growth of superconductor-semiconductor hybrid structures on silicon~\cite{HertelSilicon}. Coupling quantum devices on III-V substrates to high-quality microwave components via a flip-chip approach~\cite{FlipChip1,FlipChip2,FlipChip3,FlipChip4,FlipChip5,FlipChip6} has not yet been demonstrated. Owing to the promise of scalability presented by planar systems, investigation of this platform is of particular interest for future hybrid quantum systems. Wide employment of this technique could enable a new class of experiments using microwave techniques on lossy substrates, where the realization of high-quality microwave components is otherwise challenging. 

Here, we demonstrate microwave spectroscopy of gate- and flux-tunable ABSs in a planar JJ defined in an epitaxial InAs/Al heterostructure hosting a two-dimensional electron gas (2DEG) using a flip-chip approach. Planar JJs and coplanar microwave resonators were defined on different chips and inductively coupled via a vacuum gap. We performed single- and two-tone spectroscopy measurements of the ABS spectrum and found that the flip-chip approach, yielding low-loss resonators, enabled readout of the JJ and control of the individual ABSs. Moreover, when the electrostatically-defined transport channel in the junction hosted an isolated ABS, its transmission could be tuned up to $0.98$ via electrostatic gating. 

\begin{figure*}
	\includegraphics[width=\textwidth]{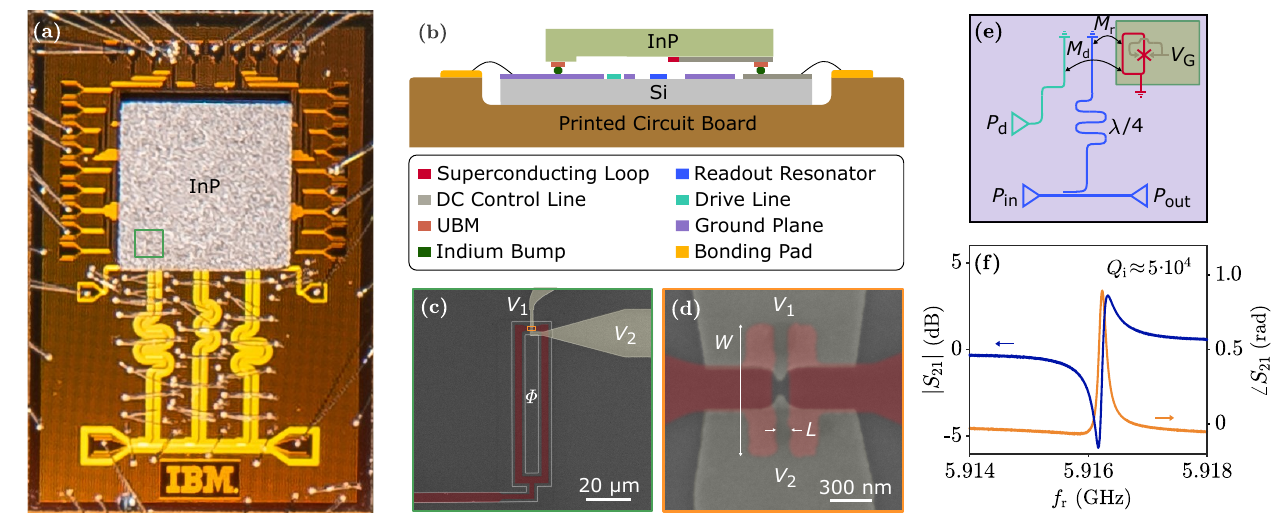}
	\caption{Overview of the flip-chip approach, device design and resonator characterization. (a) Optical image of the measured system. The Resonator Chip provided microwave components on a low-loss Si substrate. The Device Chip hosted a radio-frequency superconducting quantum interference device (rf-SQUID) on the down-facing surface of an InP chip, underneath the green box. (b) Simplified schematic cross-section of the flip-chip layout. The Resonator Chip provided coplanar waveguide (CPW) resonators capacitively coupled to a common transmission line (blue), drive lines (turquoise) and DC control lines (light gray). The two chips were galvanically coupled via indium bump bonds so that the ground of the rf-SQUID and the gates were connected to the DC control lines. (c) False-colored scanning electron micrograph of the grounded rf-SQUID (red), which was inductively coupled to a $\lambda$/4 CPW resonator and to a drive line using the flip-chip technique. (d) Gated planar Josephson junction (JJ) defined in a two-dimensional InAs/Al heterostructure, which was embedded in the rf-SQUID in (c), marked with the orange box. The Al (red) was selectively removed to form a $L\approx \mathrm{110}$~nm long and $W\approx \mathrm{940}$~nm wide JJ. A split-gate (light gray) was evaporated on top of the JJ for controllable depletion of the weak link in the exposed InAs two-dimensional electron gas (dark gray). (e) Schematic representation of the measured system, where the colors are defined in (b). (f) Magnitude (blue) and phase (orange) response of the resonator when the JJ was insulating ($\VG=-2.3$~V).}
	\label{fig1}
\end{figure*}

\section{DEVICE DESIGN, FLIP-CHIP TECHNIQUE AND RESONATOR CHARACTERIZATION}
The investigated system was composed of two chips. The Device Chip was fabricated from an InAs/Al heterostructure grown on an InP substrate~\cite{GateDependentIC}. The Resonator Chip was realized with a high-resistivity Si substrate. After the two chips were independently fabricated, they were coupled via flip-chip bonding. An optical image of the measured system, where the Device Chip is shown face-down on top of the Resonator Chip, is shown in Fig.~\ref{fig1}(a) and a simplified cross-section of the flip-chip layout is shown in Fig.~\ref{fig1}(b). 

Scanning electron micrographs of the Device Chip are shown in Figs.~\ref{fig1}(c) and (d). The active part of the Device Chip consists of a planar InAs JJ with epitaxial Al leads (red), embedded in a radio-frequency superconducting quantum interference device (rf-SQUID). The JJ had a length $L\approx\mathrm{110}~\mathrm{nm}$ and a width $W\approx\mathrm{940}~\mathrm{nm}$. Both the JJ and the rf-SQUID were obtained by wet etching of the heterostructure and selective removal of the Al. Two Ti/Au gates (light gray) with a minimal separation of $90~\mathrm{nm}$ were evaporated on top of an insulating $\mathrm{HfO_2}$ layer, and used to electrostatically define a tunable constriction in the InAs. Furthermore, the gates were arranged to fully cover InAs areas between the mesa edges and the junction, suppressing parallel conductance paths. The phase difference $\varphi$ across the JJ was tuned by applying a magnetic flux $\it{\Phi}$ through the superconducting loop. 

\begin{figure}
	\includegraphics[width=\columnwidth]{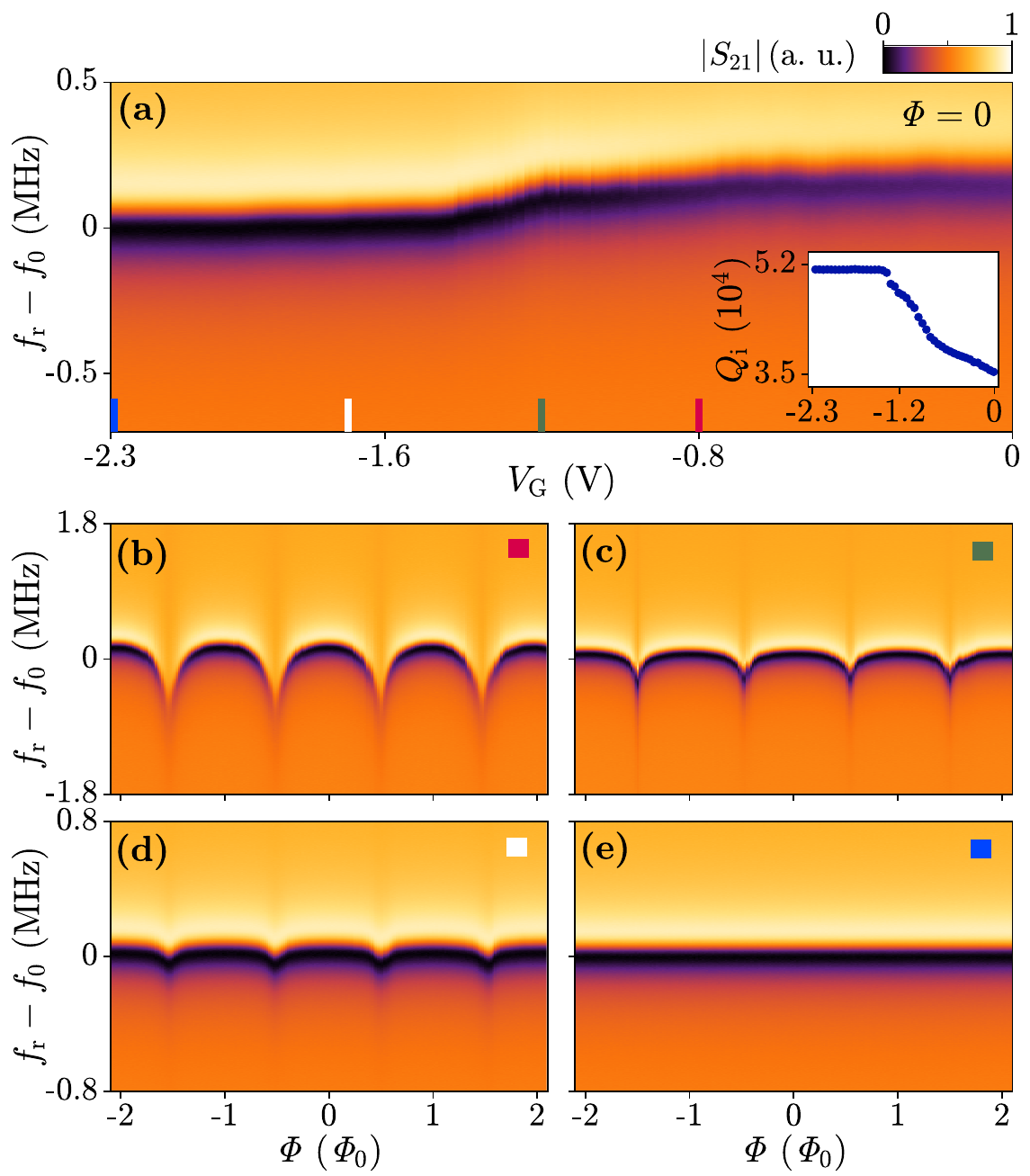}
	\caption{Single-tone spectroscopy measurements. (a) Magnitude $\abs{S_{21}}$ of the resonator transmission as a function of offset frequency $f_\mathrm{r}-f_\mathrm{0}$ and gate voltage $\VG$ at $\it{\Phi}=\mathrm{0}$. The strongest modulation of the resonance frequency $f_\mathrm{res.} \equiv \fr(\abs{S_{21}}=\mathrm{0})$ was for gate voltages between $-1.2$~V and $-1.4$~V, while for more positive and more negative gate voltages the resonance frequency was largely unaffected. The inset shows the internal quality factor $\Qi$ of the resonator as a function of $\VG$. (b-e) Flux-dependence of $\abs{S_{21}}$ at selected gate voltage values, marked with colored bars in (a). The amplitude of the modulation decreased with more negative gate voltages, until the Josephson junction was pinched-off at $\VG=-2.3$~V in panel~(e).}
	\label{fig2}
\end{figure}

The Resonator Chip was fabricated by sputtering of $200~\mathrm{nm}$ Nb and dry etching. Key elements on the Resonator Chip are represented in Fig.~\ref{fig1}(e) and included three $\lambda/4$ CPW resonators capacitively coupled to a common transmission line (blue), drive lines (turquoise), a ground plane (purple) and DC control lines (light gray). Device Chip and Resonator Chip were connected via several indium bumps, evaporated on the Resonator Chip. The bumps provided connection to the DC control lines, a ground to the rf-SQUID via under-bump metallization (UBM), and ensured mechanical stability of the structure. Particular care was taken to align rf-SQUIDs with the shorted end of the $\lambda/4$ resonators and with the drive lines to ensure the couplings with inductances $M_\mathrm{r}$ and $M_\mathrm{d}$, respectively. The green rectangle in Fig.~\ref{fig1}(b) indicates the position of the rf-SQUID measured in this work. Our flip-chip procedure allowed an alignment accuracy below $1~\mathrm{\mu m}$. The separation between the two chips was $5~\mathrm{\mu m}$, and was controlled by the size of the indium bumps and the force used during the flip-chip bonding process. To minimize dielectric losses, the Device Chip was considerably smaller than the Resonator Chip and was placed to minimize its overlap with the resonators.

The resonators were measured  with a vector network analyzer by sending a signal $\it{P}_\mathrm{in}$ into the left port of the transmission line and reading out the response $\it{P}_\mathrm{out}$ at the right port. Here, the leftmost drive line in Fig.~\ref{fig1}(b) was used to drive ABS pair transitions with a continuous microwave signal $\it{P}_\mathrm{d}$. The parasitic coupling between resonator and drive line is also reported [see Fig.~S.2 in the Supplemental Material (SM)]. The two gates were controlled by voltages $V_1$ and $V_2$, which could be set individually (see Fig.~S.5 in SM). However, a gate voltage $\VG \equiv V_1=V_2$ was applied in the remainder of this article. The magnetic flux in the rf-SQUID was produced via a home-made superconducting coil mounted on top of the printed circuit board that hosted the device.

\section{OPERATION AND CONTROLLABILITY}
Figure~\ref{fig1}(f) displays the magnitude $\abs{S_{21}}$ (blue) and phase $\angle S_{21}$ (orange) of the readout resonator while the JJ was pinched-off ($\VG=-2.3$~V). In this configuration, a bare resonator frequency $f_\mathrm{0} \approx5.9162~\mathrm{GHz}$ was measured. The asymmetry of the resonance dip in $\abs{S_{21}}$ is explained by an impedance mismatch in the circuit~\cite{AsymmetricResonator}. The internal quality factor was extracted using the circle fit technique with diameter correction~\cite{ResonatorFitting}, yielding internal and external quality factors being $\Qi \approx 5\cdot 10^4$ and $\it{Q}_\mathrm{e} \approx \mathrm{4\cdot 10^4}$, respectively. Similar values of quality factors were found for all three resonators on the Resonator Chip (see Fig.~S.3 in SM).

Figure~\ref{fig2} shows the magnitude of the complex scattering parameter $\abs{S_{21}}$ in the vicinity of $f_\mathrm{0}$ as a function of gate voltage $\VG$ and magnetic flux $\it{\Phi}$ treading the rf-SQUID. The effect on $\abs{S_{21}}$ is mediated by the variable Josephson inductance of the JJ~\cite{Haller}. Figure~\ref{fig2}(a) displays $\abs{S_{21}}$ as a function of $\VG$ measured at $\it{\Phi}=\mathrm{0}$. A clear deviation of the resonance frequency $f_\mathrm{res.} \equiv \fr(\abs{S_{21}}=\mathrm{0})$ from $f_\mathrm{0}$ appeared for less negative values of $\VG$. The inset shows the internal quality factor $\Qi$ of the resonator, extracted from the data in Fig.~\ref{fig2}(a). The flux dependence of $\abs{S_{21}}$ for four selected values of $\VG$, indicated by the colored bars in Fig.~\ref{fig2}(a), are shown in Fig.~\ref{fig2}(b-e). Periodic modulations of $f_\mathrm{res.}$ as a function of $\it{\Phi}$ were observed, with maxima and minima of $f_\mathrm{res.}$ assigned to integer and half-integer multiples of the flux quantum $\it{\Phi}_\mathrm{0}$, respectively. At $\VG=-0.8$~V, $f_\mathrm{res.}$ was modulated over all flux values $\it{\Phi}$ [Fig.~\ref{fig2}(b)]. Setting the gate more negative, the effect of $\it{\Phi}$ on $f_\mathrm{res.}$ was limited to half-integer flux quanta [Fig.~\ref{fig2}(d)]. 

Measurements presented in Fig.~\ref{fig2} are consistent with gradual depletion of lateral ABSs in the JJ~\cite{LateralABS}. For $\VG>-1.4~\mathrm{V}$, a high density of ABSs was present within the full width of the JJ. In such a situation, a quasi-continuous distribution of ABSs populates the JJ~\cite{ABSSpectrumDiffusiveLongJJ}, including several ABSs with low and high transmission probabilities~\cite{Dorokhov}. Therefore, the readout signal with frequency $\fr$ could excite transitions between ABSs at $\it{\Phi}=\mathrm{0}$, resulting in a shift of $f_\mathrm{res.}$ and a decrease in $\Qi$. When setting $\VG$ more negative, a narrow constriction is defined in the InAs, allowing the presence of only a few ABSs orthogonal to the leads. In this scenario, transitions between ABSs enabled by microwave driving are expected only in the vicinity of half-integer multiples of $\it{\Phi}_\mathrm{0}$~\cite{BeenakkerABSFormula,BagwellABSFormula}. A more quantitative description of the situation and analysis of the modulation of $f_\mathrm{res.}$, which follows Ref.~\onlinecite{Haller}, are presented in SM and further support the interpretation above. 

\section{ANDREEV BOUND STATE SPECTROSCOPY}
\begin{figure}
	\includegraphics[width=\columnwidth]{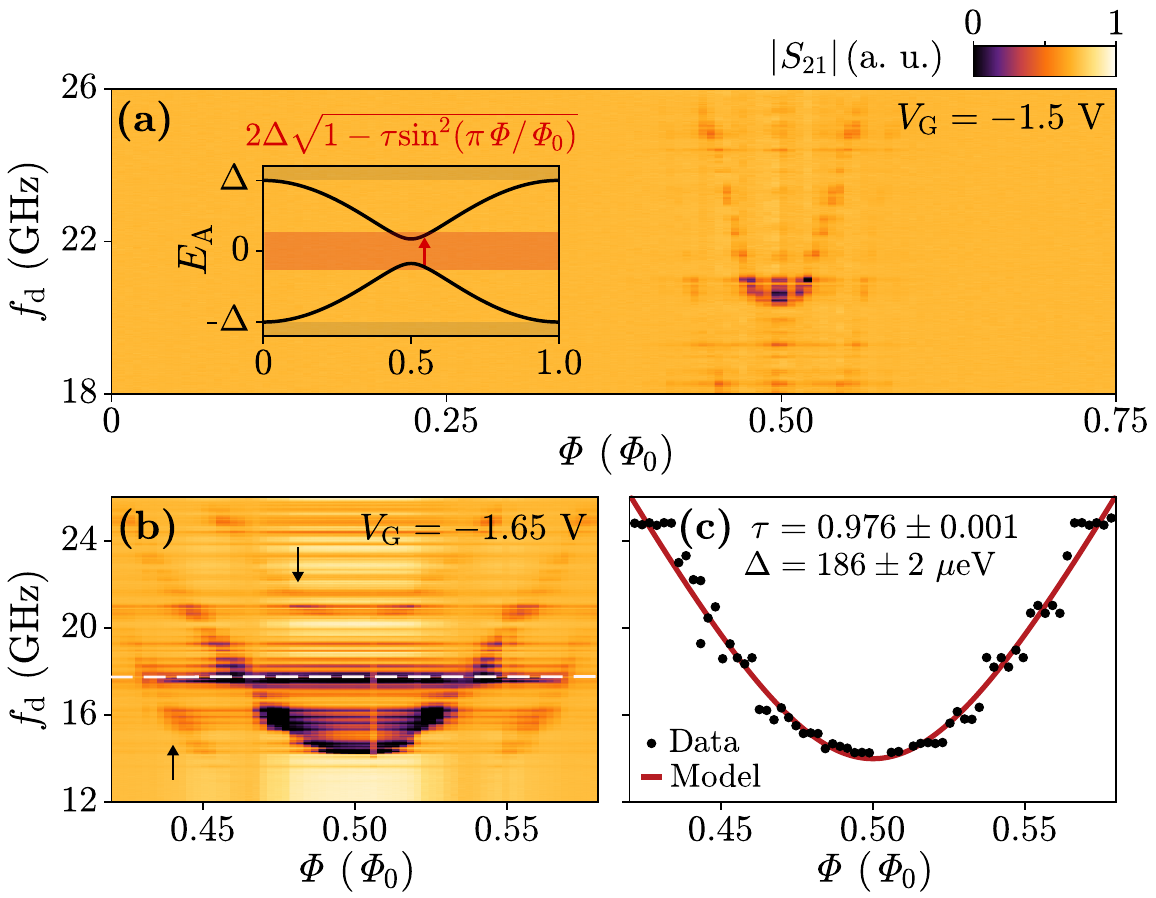}
	\caption{Two-tone spectroscopy of Andreev bound states (ABSs) as a function of magnetic flux. (a) Magnitude $\abs{S_{21}}$ of the resonator transmission in the presence of a drive tone $\it{P}_\mathrm{d}=\mathrm{-33}$~dBm with frequency $\fd$ at $V_\mathrm{G}=-1.5$~V as a function of applied flux $\it{\Phi}$. An ABS pair transition was observed around $\it{\Phi}_\mathrm{0}/\mathrm{2}$, consistent with the short junction limit. The inset shows a schematic representation of isolated ABSs in the short junction limit. The red arrow indicates an ABS pair transition and the red shaded area visualizes the accessible energy range with our experimental set-up, which was limited to 26~GHz. (b) The spectrum of ABS at $V_\mathrm{G}=-1.65$ V using $\it{P}_\mathrm{d}=\mathrm{-20}$~dBm. Aside from the ABS pair transition, weaker transitions were observed as indicated by arrows. The white dashed line indicates the second harmonic of the resonator. (c) A fit of the expression in (a) to the data extracted from the ABS pair transition in (b) yielded the transmission $\tau=0.976\pm0.001$ and the induced superconducting gap $\Delta=186\pm2$~$\mu$eV of the observed ABS.}
	\label{fig3}
\end{figure}

A direct visualization of the ABSs was obtained via two-tone spectroscopy measurements as a function of $\it{\Phi}$, as presented in Fig.~\ref{fig3}(a). For these measurements, the JJ was tuned to the few-modes regime via $\VG$ and a continuous drive tone of variable frequency $\fd$ was applied to the drive line while monitoring magnitude and phase of $S_{21}$ at $\fr$ \footnote{The frequency $\fr$ had to be determined for each value of $\it{\Phi}$}. In this situation, the drive tone excited transitions between ABS pairs within the superconducting gap, as schematically shown in the inset of Fig.~\ref{fig3}(a). For a short, single-channel weak link the energy of the ABSs is given by:
\begin{equation}
	E_\mathrm{A}=\pm \Delta \sqrt{1-\tau \sin^2 (\mathrm{\pi} \it{\Phi}/ \it{\Phi}_\mathrm{0})},
	\label{eq1}
\end{equation}
where $E_\mathrm{A}$ is the ABS energy, $\Delta$ is the induced superconducting gap and $\tau$ is the transmission of the ABS~\cite{BeenakkerABSFormula,BagwellABSFormula}. The energy needed for an ABS pair transition is $\mathrm{2}E_\mathrm{A}$. Red shading indicates the experimentally accessible energy range, which intersects the ABS spectrum close to $\it{\Phi} = \it{\Phi}_\mathrm{0}/\mathrm{2}$.
As expected, ABSs pair transitions were observed only in the vicinity of $\it{\Phi} = \it{\Phi}_\mathrm{0}/\mathrm{2}$. Accordingly, this two-tone spectroscopy measurement supports the interpretation that the planar JJ was in the ballistic short junction regime. 

At $\VG=-1.65$~V, an ABS pair transition with a minimum frequency of 14~GHz was probed [Fig.~\ref{fig3}(b)]. In addition to the pair transition, weaker features at lower and higher $\fd$ than the main ABS transition were observed, as indicated by arrows. These replicas, which are spaced by $f_\mathrm{res.}$ from the ABS transition, are explained by finite photon population in the resonator during ABS drive. In this context, additional lines appear when the resonator emits or absorbs the additional photon that satisfies the energy relation $2E_\mathrm{A} \pm hf_\mathrm{res.} = h\fd$, where $\it{h}$ is Planck's constant~\cite{FlipChip4}. 
The horizontal resonances in Fig.~\ref{fig3}(b) are associated to standing waves in the microwave cabling of the measurement circuit. Such flux-independent resonances were previously observed in two-tone spectroscopy measurements in the presence of strong driving signals and are likely caused by impedance mismatch along the signal propagation path~\cite{FlipChip4, NanowireLongJunction6}. In particular, the strongest resonance, marked with a white dashed line, corresponds to the second harmonic of the resonator. All of these lines became more pronounced at higher driving powers (see Fig.~S.11 in SM).

The measured ABS pair transition was extracted from Fig.~\ref{fig3}(b) and fitted with $2E_\mathrm{A}$, with $E_\mathrm{A}$ from Eq.~\ref{eq1} to determine the induced superconducting gap $\Delta$ and the transmission $\tau$ of the ABS. From the fit, shown in Fig.~\ref{fig3}(c), $\Delta = 186 \pm 2~\mu$eV and $\tau = 0.976 \pm 0.001$ were obtained. The value of $\Delta$ is consistent with previous transport studies of hybrid superconductor-semiconductor heterostructures~\cite{InducedGapAl2DEG1,InducedGapAl2DEG2}. Following Ref.~\onlinecite{AtomicContact2}, we estimated the coupling strength $\it{g(\Phi,\tau)}$ between the isolated ABS and the resonator. At $\it{\Phi} = \it{\Phi}_\mathrm{0}/\mathrm{2}$ a coupling strength of $\it{g} \approx \mathrm{35}$~MHz was extracted. 

\section{ELECTROSTATIC TUNING OF ISOLATED ANDREEV BOUND STATES}
\begin{figure}
	\includegraphics[width=\columnwidth]{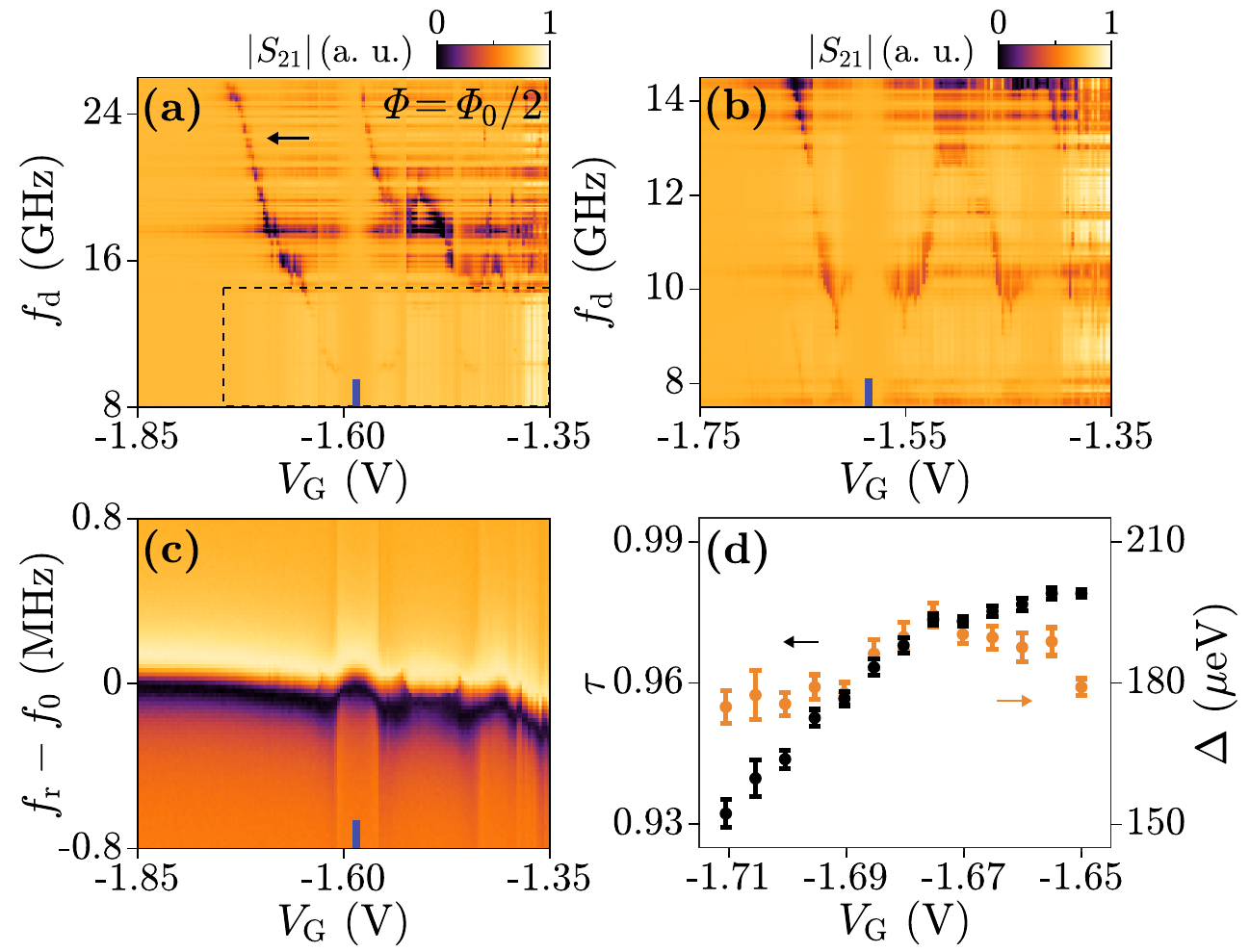}
	\caption{Two-tone spectroscopy of Andreev bound states (ABSs) as a function of gate voltage. (a) Magnitude $\abs{S_{21}}$ of the resonator transmission in the presence of a drive tone $\it{P}_\mathrm{d}=\mathrm{-24}$~dBm with frequency $\fd$ at $\it{\Phi} = \it{\Phi}_\mathrm{0}/\mathrm{2}$ as a function of gate voltage $\VG$. Isolated ABS pair transitions were observed from $\VG= -1.4$~V to $\VG=-1.75$ V. At more positive gate voltages, ABS transitions could no longer be resolved. (b) Zoomed-in measurement of the region marked with a box in (a) using higher drive power, to enhance the visibility of the ABS transition at lower frequencies. An accidental resonance in the junction decreased the ABS frequency in the vicinity of  $\VG=-1.6$~V (blue marker). (c) Single-tone spectroscopy measurement corresponding to the gate voltage range in (a). (d) Transmission~$\tau$ and induced superconducting gap~$\Delta$ extracted from the ABS pair transition, indicated by the arrow in (a) as a function of gate voltage $\VG$.} 
	\label{fig4}
\end{figure}

To further investigate the ABS spectrum in the planar JJ, the applied flux was fixed to $\it{\Phi} = \it{\Phi}_\mathrm{0}/\mathrm{2}$ and two-tone spectroscopy measurements were performed as a function of the gate voltage $\VG$, as shown in Fig.~\ref{fig4}(a). No isolated ABS pair transitions could be resolved for $\VG > -1.4$~V, indicating a dense spectrum of ABSs in the JJ. This is consistent with the observations from single-tone measurements. At more negative $\VG$, two isolated ABS pair transitions were observed. Their transition frequencies increased with decreasing $\VG$ until, below $\VG = -1.75$~V, no ABS transitions were observed in the experimentally accessible frequency range. 

Figure \ref{fig4}(b) depicts a zoom-in of the region marked with a box in Fig.~\ref{fig4}(a), measured at higher drive power to increase the visibility of ABS transitions for~$\fd<$14~GHz. The frequency of the isolated ABS, indicated with an arrow in panel~(a), continuously decreased down to $\fd~=~10$~GHz towards $\VG = -1.6$~V. The pair transition signal disappeared in the vicinity of $\VG=-1.6$~V (blue marker) before its frequency increased again. This frequency modulation of the ABS pair transition is explained by the formation of an accidental, incoherent resonance in the junction. Such resonances were previously shown to enhance the transmission $\tau$ of the JJ~\cite{AndreevQD1,AndreevQD2,Resonator2DEG2}. The single-tone measurement of the resonator corresponding to the gate range in Fig.~\ref{fig4}(a) is depicted in Fig.~\ref{fig4}(c). Here, the incoherent resonance manifested itself as a local increase in the resonator frequency at $\VG=-1.6$~V (blue marker). Additional flux-dependent single- and two-tone spectroscopy measurements in the vicinity of the incoherent resonance are depicted in Fig.~S.12 in SM. 

Flux-dependent two-tone spectroscopy measurements of the isolated ABS, indicated with an arrow in Fig.~\ref{fig4}(a), were performed at multiple gate voltages (see Fig.~S.13 in SM). For each data set, $\Delta$ and $\tau$ were extracted as in Fig.~\ref{fig3}(c). The results are summarized in Fig.~\ref{fig4}(d). The induced superconducting gap was nearly constant with an average value of 184~$\mu$eV, while the transmission was found to increase continuously with gate voltage, approaching $\tau=0.98$ at $\VG=-1.65$~V. The limited spectroscopic window and the absence of reliable flux-dependent two-tone spectroscopy measurements above $\VG=-1.65$~V prevented the probing of a larger gate voltage range. Nevertheless, the access and control of an isolated ABS was successfully demonstrated.

\section{CONCLUSION}
We presented a framework for microwave spectroscopy measurements in two-dimensional heterostructures using a flip-chip approach. We fabricated a planar JJ where the conductive channel was defined by electrostatic gating, allowing the probing of an isolated ABS with transmission up to $0.98$. Efficient and reliable coupling between superconductor-semiconductor hybrid devices and microwave electronics fulfills a crucial requirement towards scalable architectures for Andreev qubits, Andreev molecules and topological phenomena.

\begin{acknowledgments}
We thank the Cleanroom Operations Team of the Binnig and Rohrer Nanotechnology Center (BRNC) for their help and support. We also thank Marcelo Goffman, Hugues Pothier and Cristi\'{a}n Urbina for useful comments and questions. F.~N. acknowledges support from the European Research Council (grant number 804273) and the Swiss National Science Foundation (grant number 200021\_201082).
\end{acknowledgments}

\section*{data availability}
Data presented in this work will be available on Zenodo. The data that support the finding of this study are available upon reasonable request from the corresponding author.

\bibliography{Bibliography.bib}

\newpage
\setcounter{section}{0}
\onecolumngrid

\newcounter{myc}
\renewcommand{\thefigure}{S.\arabic{myc}}

\newcounter{mye}
\renewcommand{\theequation}{S.\arabic{mye}}

\newpage	
\setlength{\parskip}{0pt} 

\section*{Supplemental Material: Flip-chip-based microwave spectroscopy of Andreev bound states in a planar Josephson junction}

\section{Material and fabrication}
Resonator and Device Chips were made separately and subsequently flip-chip bonded using indium bumps. The Device Chip was fabricated from a heterostructure grown on an InP (001) substrate using molecular beam epitaxy techniques. The heterostructure consisted of an $\mathrm{In_xAl_{1-x}As}$ buffer and a 8~nm thick InAs quantum well, confined by $\mathrm{In_{0.75}Ga_{0.23}As}$ barriers 13~nm below the surface. A 10~nm thick Al layer was deposited on top of the heterostructure, in the same chamber as the III-V growth without breaking the vacuum. From gated Hall bar measurements, after removal of the Al, the wafer had a peak mobility of $\mu =$~18000~$\mathrm{cm^2 V^{-1} s^{-1}}$ and an electron density of $n = 8 \cdot 10^{11}~\mathrm{cm^{-2}}$, resulting in an elastic mean free path $l_\mathrm{e} = \mu \hbar \sqrt{2 \mathrm{\pi} n}/\mathrm{e} \approx$ 270~nm. The superconducting coherence length was estimated as $\xi=\hbar v_\mathrm{F}/(\pi \Delta) \approx$ 1.3~$\mu \mathrm{m}$ for a ballistic junction and $\xi=\sqrt{\hbar D/\Delta} \approx$~740~nm for a diffusive junction, with diffusion coefficient $D=v_\mathrm{F}l_\mathrm{e}/2$, Fermi velocity $v_\mathrm{F}=\hbar \sqrt{2\pi n}/m^*$ and induced superconducting gap $\Delta = 184~\mu \mathrm{eV}$ determined from two-tone spectroscopy measurements. The sample was defined by first isolating the mesa structures by selectively removing the top Al layer with Transene D etchant before etching $\approx$~290 nm into the III-V heterostructure using a chemical wet etch solution of $\mathrm{H_2O : C_6H_8O_7 : H_3PO_4 : H_2O_2}$ (220 : 55 : 3 : 3). The under bump metalization (UBM) was deposited on top of the Al contacts using lift-off techniques. The $L\approx110$~nm long and $W\approx940$~nm wide Josephson junction~(JJ) was patterned on top of the mesa, by selective removal of the Al with Transene D etchant. Then a 15~nm layer of $\mathrm{HfO_2}$ was deposited using atomic layer deposition technique, before evaporating metallic gate electrodes. In the region where the resonators are close to the Device Chip, the InAlAs buffer was etched away. As a last step, the $\mathrm{HfO_2}$ on top of the UBM was removed to have a galvanic connection between the rf-SQUID on the Device Chip and the DC control lines on the Resonator Chip. 

The Resonator Chip was fabricated from a high resistivity ($\rho~\mathrm{>~10000~\Omega \cdot}$cm) intrinsic Si $\mathrm{\langle100\rangle}$ wafer. After removing the native oxide using hydrofluoric acid, 200 nm of Nb was sputtered on the wafer. In a single lithographic step, readout resonators, device drive and DC control lines, as well as holes in the ground plane, were patterned into the Nb with a $\mathrm{Cl_2/Ar}$ dry etching procedure. Afterwards, the In was evaporated directly on top of the Nb using lift-off techniques. Hydrofluoric acid was used again to remove the surface oxide of the Nb and In before flip-chip bonding. 

The Device Chip was bonded to the Resonator Chip with a flip-chip bonder (Karl Suss FC 150). To facilitate the alignment of both chips, alignment markers were defined during the mesa etch and resonator etch step. After gluing the sample into the printed circuit board (PCB) holder, the microwave and DC lines from the PCB were connected to the corresponding bonding pads on the Resonator Chip with Al bond wires. Moreover, Al air bridges across the transmission line, resonators and drive lines connecting the different ground planes ensured a homogeneous ground potential all over the Resonator Chip.

\section{Experimental setup}
\begin{figure}[H]
\begin{center}
	\centering
	\includegraphics[width=\columnwidth]{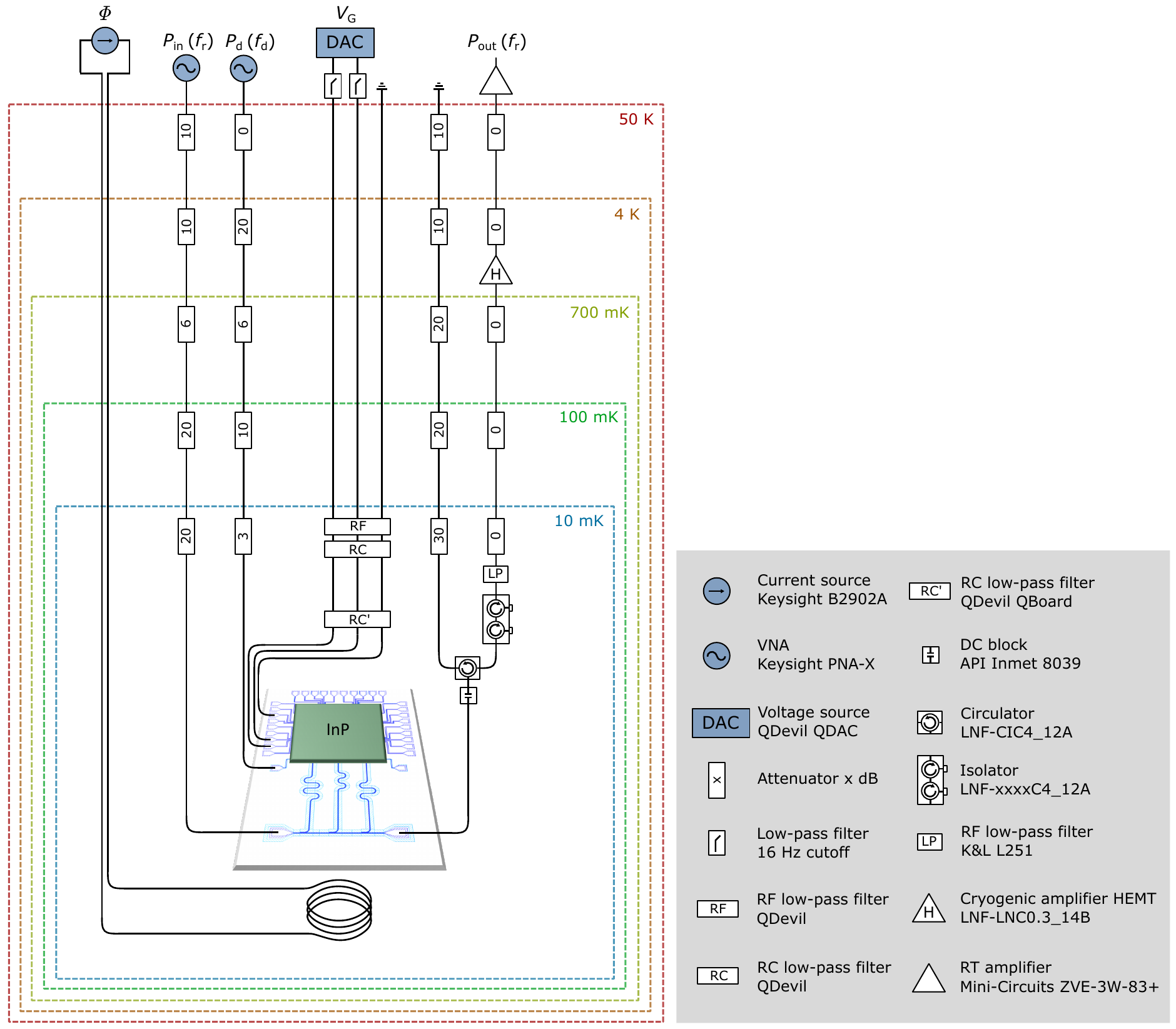}
	\caption{Schematic of the measurement setup. The resonator readout and device drive tones with frequencies $f_\mathrm{r}$ and $f_\mathrm{d}$ were applied using a vector network analyzer (VNA), corresponding to $P_\mathrm{in}$ and $P_\mathrm{d}$, respectively. After 66 dB and 39 dB attenuation of $P_\mathrm{in}$ and $P_\mathrm{d}$ at different temperature stages of the dilution refrigerator, the signals reached the Resonator Chip. The resonator transmission signal passed through a DC block followed by a circulator, a dual isolator and an RF low-pass filter. The signal was amplified with a cryogenic HEMT amplifier at 4~K and a room-temperature amplifier before being detected by the VNA ($P_\mathrm{out}$). An extra RF line, which could be used for reflectometry measurements, stayed grounded during the experiments. The gates were controlled with DAC DC voltage sources. The DC lines were filtered using a homemade low-pass filter at room temperature, QDevil RF and RC filters at base temperature and an RC filter on the printed circuit board. The external flux $\it{\Phi}$ was supplied by sourcing a DC current through a coil, mounted on top of the sample space.}
	\label{figS1}
\end{center}
\end{figure}

The measurements were carried out in a BlueFors BF-LD cryogen-free dilution refrigerator with a mixing chamber base temperature of 9~mK. A schematic of the wiring setup is depicted in Fig.~\ref{figS1}. The resonator readout and device drive tones $P_\mathrm{in}$ and $P_\mathrm{d}$ were continuously applied using a Keysight PNA-X vector network analyser (VNA). Signals were attenuated in the lines by 66~dB and 39~dB, respectively for readout and drive tones. Readout was performed in transmission mode, where the signal from the transmission line on the Resonator Chip, after interacting with the device, passed through a DC block, a circulator, a dual isolator and an RF low-pass filter and was subsequently amplified with a cryogenic high electron mobility (HEMT) amplifier (bandwidth 0.3-14~GHz, average gain 37~dB) followed by a room-temperature amplifier (bandwidth 2-8~GHz, average gain 35~dB) before reaching the VNA again ($P_\mathrm{out}$). The voltages on gates $V_1$ and $V_2$ were controlled with DC voltages sourced with a QDevil QDAC. The voltages applied via the DC loom were filtered with QDevil RF and RC filters (cutoff above 80~MHz and 65~kHz, respectively) before reaching the Device Chip. The chips were mounted on a QDevil high frequency PCB with no additional shielding present. A vector magnet was present in the dilution refrigerator but not used in this study and hence left grounded. Instead, the flux was supplied by sourcing a current through a coil, mounted in the near vicinity of the chips.

\raggedbottom
\newpage

\setcounter{myc}{1}

\newpage

\section{Resonator Chip characterization}
\begin{figure}[H]
	\centering
	\includegraphics[width=0.71\columnwidth]{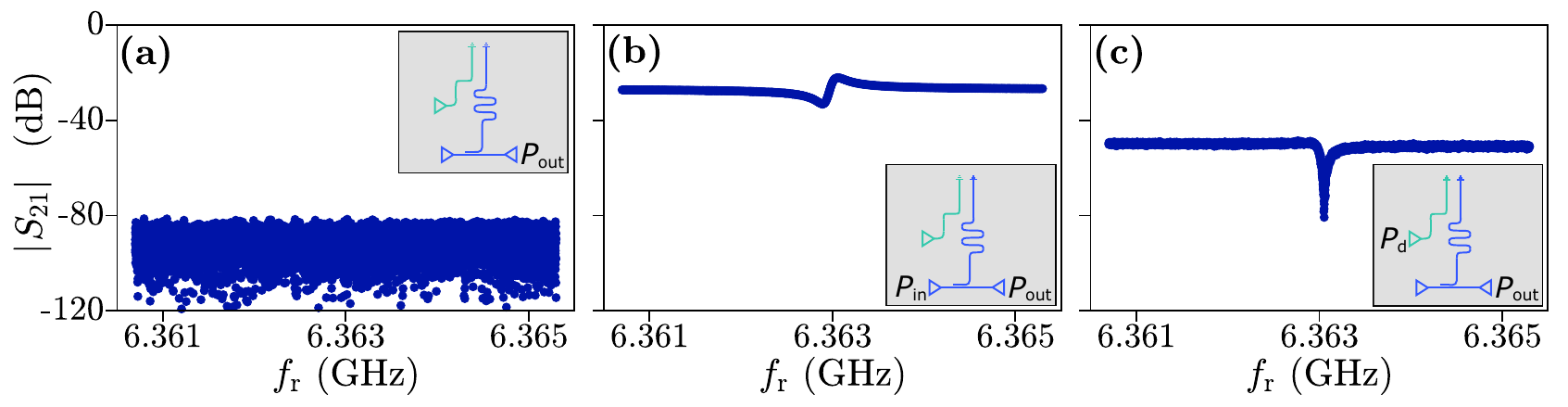}
	\caption{Characterization of the unintentional coupling between device drive line and readout resonator. (a) A background level of $-80$~dB ($P_\mathrm{in} = -30$~dBm) was measured by applying a signal with frequency $f_\mathrm{r}$ to an RF line, which was not connected to the printed circuit board, and monitoring the magnitude of transmission $\abs{S_\mathrm{21}}$ by detecting the output power $P_\mathrm{out}$ emerging from the transmission line using a vector network analyzer (VNA). The frequency $f_\mathrm{r}$ was swept across the resonance frequency ($f_\mathrm{res.}=6.363$~GHz) of the rightmost resonator on the Resonator Chip. The background level was independent of $f_\mathrm{r}$ and consistent with the noise floor of the VNA. (b) The signal with power $P_\mathrm{in} = -30$~dBm was applied to the left port and $P_\mathrm{out}$ was measured at the right port of the transmission line. The magnitude of transmission $\abs{S_\mathrm{21}}$ was monitored as a function of frequency $f_\mathrm{r}$. In this case, the background was at $-25$~dB with a pronounced resonance feature at $f_\mathrm{r}=6.363$~GHz consistent with $f_\mathrm{res.}$. (c) A signal with power $P_\mathrm{d} = -30$~dBm was applied to the device drive line and $P_\mathrm{out}$ was measured at the right port of the transmission line. Here, the background was at $-50$~dB indicating an unintentional coupling with $-25$~dB attenuation between device drive line and readout resonator.}
	\label{figS2}
\end{figure}

The unintentional coupling between the drive line and the resonator was characterized by performing a series of single-tone spectroscopy measurements in the absence of Device Chip, as depicted in Fig.~\ref{figS2}. The frequency $f_\mathrm{r}$ of the signal was swept across the resonance frequency ($f_\mathrm{res.}=6.363$~GHz) of the rightmost resonator in Fig.~1(a) of the Main Text. In panel~(a), a background level of $-80$~dB ($P_\mathrm{in} = -30$~dBm) was measured by applying a signal with frequency $f_\mathrm{r}$ to an RF line, which was not connected to the PCB, and monitoring the magnitude of transmission  $\abs{S_\mathrm{21}}$ by detecting the output power $P_\mathrm{out}$ using a VNA. The background level was independent of $f_\mathrm{r}$ and consistent with the noise floor of the VNA. In (b), the magnitude of transmission $\abs{S_\mathrm{21}}$ is shown when a signal with power $P_\mathrm{in} = -30$~dBm was applied to the left port of the transmission line and measured at its right port. The readout frequency $f_\mathrm{r}$ was swept across $f_\mathrm{res.}$ and a background level of $-25$~dB was observed. Panel~(c) shows $\abs{S_\mathrm{21}}$ when the signal with power $P_\mathrm{d} = -30$~dBm was applied to the drive line and $P_\mathrm{out}$ was measured at the right port of the transmission line. In this configuration, the background was at $-50$~dB, which is higher than the background level of $-80$~dB, indicating an unintentional coupling between the drive line and the resonator. The $-25$~dB attenuation between device drive line and resonator was a trade-off between sufficiently strong coupling between device drive line and rf-SQUID and minimization of the contribution of losses to the resonator. However, this drive line did not limit the quality factor of the resonator, instead, the Device Chip was the dominant component of the loss tangent. 

\setcounter{myc}{2}

\begin{figure}[H]
	\centering
	\includegraphics[width=0.56\columnwidth]{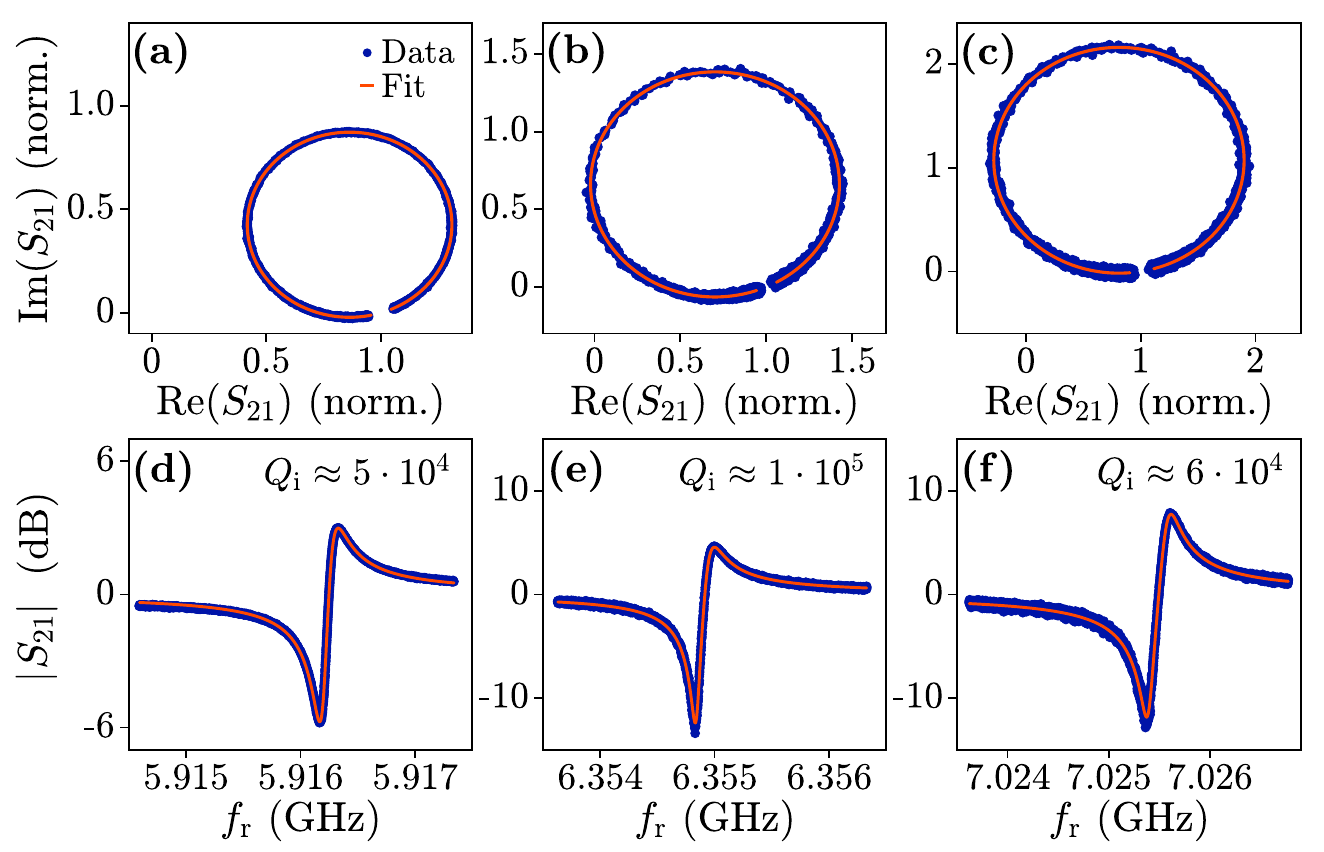}
	\caption{Resonator characterization. (a-c) For all three resonators on the Resonator Chip in Main Text Fig.~1(a), real $\mathrm{Re}(S_\mathrm{21})$ and imaginary $\mathrm{Im}(S_\mathrm{21})$ part of the transmission $S_\mathrm{21}$ (blue) is plotted together with a fit (red) to determine the internal quality factor $\Qi$ of the resonators. The corresponding magnitude response $\abs{S_\mathrm{21}}$ as a function of readout frequency $f_\mathrm{r}$ is displayed in (d-f). Panels (a) and (d) correspond to the resonator used in the measurements of the Main Text, while panels (b) and (e) [(c) and (f)] correspond to the center [rightmost] resonator in Fig.~1(a) of the Main Text. The measurements were performed while the corresponding Josephson junction was pinched-off. The circle fit technique with diameter correction from Ref.~\cite{ResonatorFitting} was used to extract the internal quality factors $\Qi$ of the resonators, yielding values between $5\cdot 10^4$ and $1\cdot 10^5$ as shown in (d-f).}
	\label{figS3}
\end{figure}

The three resonators on the Resonator Chip in Main Text Fig.~1(a) were characterized. The results are summarized in Fig.~\ref{figS3}. The measurements were performed while the corresponding JJ on the Device Chip was fully depleted. The circle fit technique with diameter correction from Ref.~\cite{ResonatorFitting} was used to extract the internal quality factor $\Qi$ of the resonators. Values of $\Qi$ ($\Qe$) between $5\cdot 10^4$ ($1.8\cdot 10^4$) and $1\cdot 10^5$ ($4.1\cdot 10^4$) were obtained. In comparison to previous work \cite{Resonator2DEG2,Resonator2DEG3,Resonator2DEG4,Resonator2DEG5}, these consistently high values of $\Qi$ demonstrate the potential of employing flip-chip techniques for microwave experiments with devices fabricated on III-V substrates. Improving materials as well as fabrication, deep etching into the silicon substrate \cite{DeepEtchResonator} or reducing the amount of III-V materials in the vicinity of the microwave components \cite{CasparisGatemon2DEG,HertelSilicon} could further enhance the quality factors of the microwave resonators.

\setcounter{myc}{3}

\section{Measurement procedure and data processing}
In single-tone spectroscopy measurements, a signal $P_\mathrm{in}$ with frequency $f_\mathrm{r}$, swept around the readout resonator frequency, was sent to the transmission line. Ater the interaction with the readout resonator, the magnitude $\abs{S_{21}}$ and phase $\angle S_{21}$ of the transmitted part of the input signal ($P_\mathrm{out}$) was measured using a VNA. To acquire flux- and gate-dependent measurements, this procedure was repeated at multiple values of $\it{\Phi}$ or $V_\mathrm{G}\equiv V_1=V_2$, respectively. We assigned $\it{\Phi}=$~0 to the central DC current value, where the resonator frequency was maximum and independent of small deviations of DC current. Since the resonator frequency is modulated with a period of one flux quantum $\it{\Phi}_\mathrm{0}$, $\it{\Phi}=$~0 was defined within the flux quantum when no current was applied to the coil.

For two-tone spectroscopy measurements, the readout frequency $f_\mathrm{r}$ was first swept around the resonance frequency of the readout resonator to determine the minimum in the magnitude $\abs{S_{21}}$ corresponding to $f_\mathrm{res.}$. The readout frequency was then offset by 80 kHz, $f_\mathrm{r}=f_\mathrm{res.}+80$~kHz, to position the readout tone on the slope of the resonator response to increase the sensitivity of shifts in $f_\mathrm{res.}$ for signal magnitude measurements. The spectroscopy measurement was then performed by monitoring the resonator response at the aforementioned fixed $f_\mathrm{r}$, while simultaneously, a continuous drive tone $P_\mathrm{d}$ at frequency $f_\mathrm{d}$ was applied to the device drive line. The drive tone with frequency $f_\mathrm{d}$ was swept while measuring the magnitude and phase of the resonator signal $P_\mathrm{out}$. Finally, the background of the resonator response was removed by subtracting the median of the $f_\mathrm{d}$ sweep. For flux- and gate-dependent two-tone spectroscopy measurements this procedure was repeated at each value of $\it{\Phi}$ or $V_\mathrm{G}$, respectively.

In all single- and two-tone spectroscopy plots, the colorbar of the magnitude $\abs{S_{21}}$ was rescaled to [0,1]. This was done by finding the minimum value of $\abs{S_{21}}$ for each dataset. Then, this minimum magnitude was added to all values of $\abs{S_{21}}$ in the dataset, such that the new dataset consisted of only positive values. Afterwards, the maximum value of $\abs{S_{21}}$ was found and all $\abs{S_{21}}$ were normalized to this value. 

The transmission $\tau$ and induced superconducting gap $\Delta$ of the JJ were obtained by fitting two-tone spectroscopy data representing a $\it{\Phi}$-dependent Andreev bound state (ABS) pair transition using Eq.~\ref{ABSTransition}. In the short junction limit, an ABS pair transition frequency $f_\mathrm{ABS}$ is described by: 
\setcounter{mye}{1}
\begin{equation}
	f_\mathrm{ABS}=2E_\mathrm{A}=2 \Delta \sqrt{1-\tau \sin^2 (\mathrm{\pi} \it{\Phi}/ \it{\Phi}_\mathrm{0})},
	\label{ABSTransition}
\end{equation}
where $E_\mathrm{A}$ is the energy of a single-channel ABS \cite{BeenakkerABSFormula,BagwellABSFormula}. First, resonances caused by the device drive line and measurement circuit were filtered out by taking the gradient along $\it{\Phi}$ for each $f_\mathrm{d}$. Then, a Sobel filter was applied to extract all edges of the color plot. To remove the weaker higher-order transitions, the maximum value of the response along the extracted edges was taken at every flux $\it{\Phi}$. The remaining data points were fitted to Eq.~\ref{ABSTransition}. Due to the spurious resonances detected in the measurement circuit, it was not possible to fit the two-tone spectrum at every flux $\it{\Phi}$ with a Lorentzian function to extract the data points corresponding to the energy of the ABS pair transition. To overcome this problem, the window function could be used, but this required assumptions about the flux-dependence of the ABS energy.

\section{Additional measurements}
\subsection{Additional single-tone spectroscopy data}
\begin{figure}[H]
	\centering
	\includegraphics[width=0.5\columnwidth]{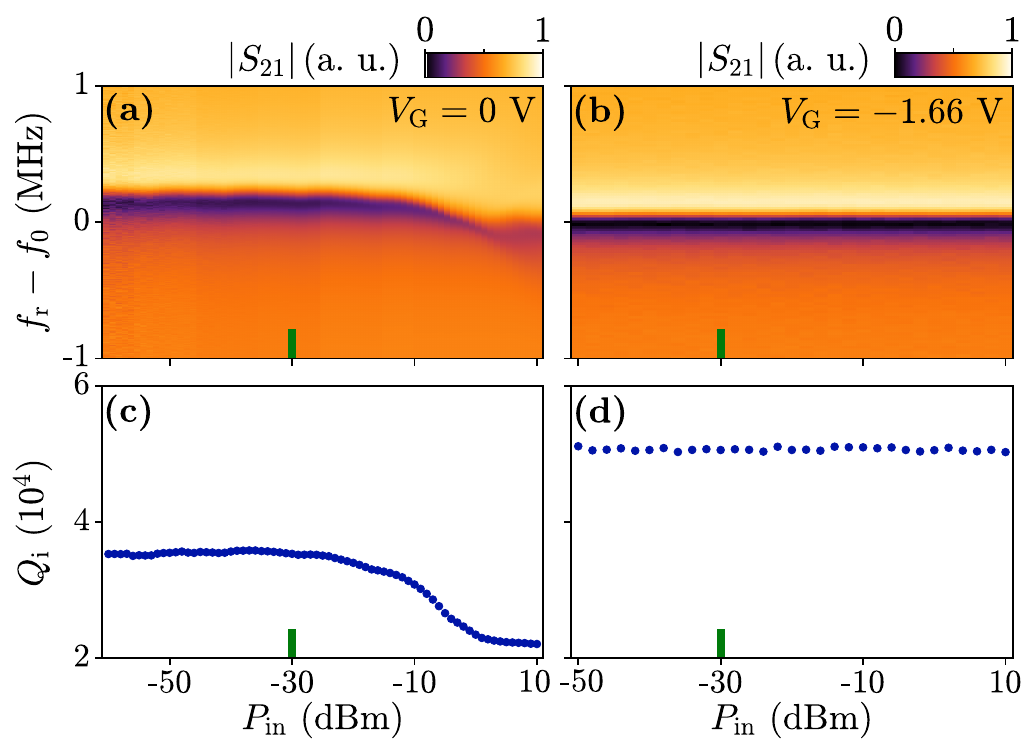}
	\caption{Resonator response as a function of readout power. (a) Magnitude $\abs{S_{21}}$ of the resonator transmission as a function of readout tone power $P_\mathrm{in}$ as well as frequency $f_\mathrm{r}$ at $\it{\Phi}=\mathrm{0}$ and $V_\mathrm{G}=0$~V. The Josephson junction (JJ) hosted a high density of Andreev bound states (ABSs), thereby shifting the resonance frequency downwards at large $P_\mathrm{in}$. (b) Same as (a) for $V_\mathrm{G}=-1.66$~V. Only few ABSs were present in the JJ. In this situation, the resonance frequency remained constant with increasing $P_\mathrm{in}$. Internal quality factors $\Qi$ as a function of $P_\mathrm{in}$ were extracted from the resonance curves in (a) and (b) and are displayed in panels~(c) and (d), respectively. The green markers at $P_\mathrm{in}=-30$~dBm. indicate the power of the resonator probe tone $P_\mathrm{in}$ used for the measurements in the Main Text}
	\label{figS4}
\end{figure}

The influence of readout power on resonator response is shown in Fig.~\ref{figS4}. For $V_\mathrm{G}=0$~V and $V_\mathrm{G}=-1.66$~V, the readout power $P_\mathrm{in}$ was swept at fixed flux $\it{\Phi}=\mathrm{0}$. The green markers in Fig.~\ref{figS4}(a) and (c) indicate the value of $P_\mathrm{in}$ at which the measurements in the Main Text were taken ($-30$~dBm). The change in resonance frequency at high readout power in Fig.~\ref{figS4}(a) is commonly observed in coupled resonator-rf-SQUID systems under strong driving, and could signify non-linear effects such as over-driving of the resonator or saturation of the available ABSs \cite{Power1,Haller}. Moreover, the internal quality factor $\Qi$ was also reduced as the readout power was increased beyond $-30$~dBm as shown in panel~(c). Yet, when the gate voltage was set to $V_\mathrm{G}=-1.66$~V, these effects were no longer present: the resonance frequency and $\Qi$ remained constant as the readout power was increased as depicted in panels~(b) and (d). Based on these results, we conclude that the change in resonance frequency and $\Qi$ were related to the density of ABSs. At $V_\mathrm{G}=-1.66$~V, no ABSs were available for microwave-induced transitions at $\it{\Phi}=\mathrm{0}$, because in this configuration the JJ was likely in the ballistic short junction regime ($L < l_\mathrm{e} \ll \xi$). At $V_\mathrm{G}=0$~V lateral ABSs, which were extended longer than $l_\mathrm{e}$ and comparable to $\xi$, were present in the JJ, yielding a minigap with many ABSs with low and high transmission probabilities~\cite{ABSSpectrumDiffusiveLongJJ,Dorokhov}. Therefore, at high readout power the number of transitions between ABSs was increased, resulting in a larger number of loss channels and thus a decrease in $\Qi$ in Fig.~\ref{figS4}(c)~\cite{Losses1}. Moreover, the transitions to ABSs at higher energy, that have lower transmission, reduced the supercurrent in the rf-SQUID, yielding an increase of the inductance of the JJ for high readout powers and thus a decrease in the resonance frequency.

\setcounter{myc}{4}

\setcounter{myc}{5}
\begin{figure}[H]
	\centering
	\includegraphics[width=0.5\columnwidth]{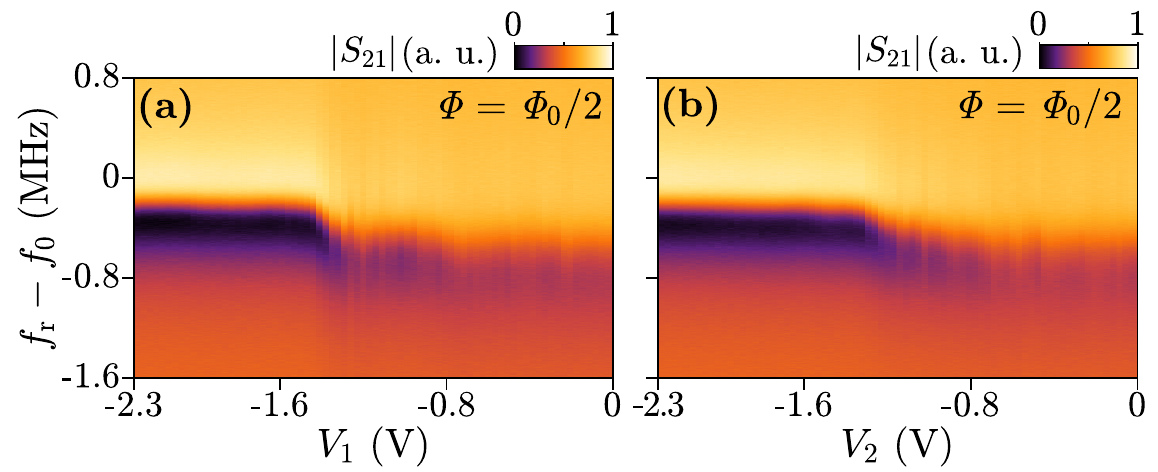}
	\caption{Resonator response at $\it{\Phi}=\it{\Phi}_\mathrm{0}/\mathrm{2}$ when tuning both gates of the Josephson junction (JJ) individually (see Fig.~1(c) of the Main Text). (a) Magnitude $\abs{S_{21}}$ was monitored around the resonance frequency of the readout resonator as a function of gate voltage $V_\mathrm{1}$, while $V_\mathrm{2}=0$~V. (b) Same as in (a), but $\abs{S_{21}}$ as a function of $V_2$ with $V_1=0$~V. For both gates, a decrease in gate voltage led the resonance frequency of the resonator to increase (after an initial plateau) before it saturated below $V_\mathrm{1,2}\approx-1.4$~V. Moreover, the JJ could be pinched-off ($f_\mathrm{r}=f_\mathrm{0}$) only if both gates were energized.}
	\label{figS5}
\end{figure}

\setcounter{myc}{6}
\begin{figure}[H]
	\centering
	\includegraphics[width=0.5\columnwidth]{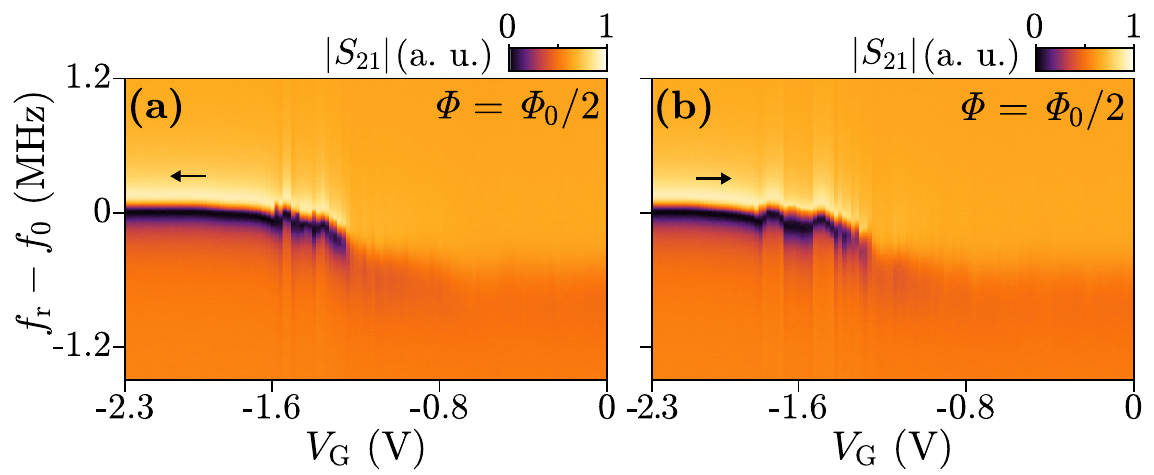}
	\caption{Characterization of gate hysteresis. Magnitude $\abs{S_{21}}$ was measured around the resonance frequency of the readout resonator while sweeping the gate voltage $V_\mathrm{G}$ (a) down and (b) up at $\it{\Phi}=\it{\Phi}_\mathrm{0}/\mathrm{2}$. The small gate hysteresis did not affect the two-tone spectroscopy experiments.}
	\label{figS6}
\end{figure}

\begin{figure}[H]
	\centering
	\includegraphics[width=0.71\columnwidth]{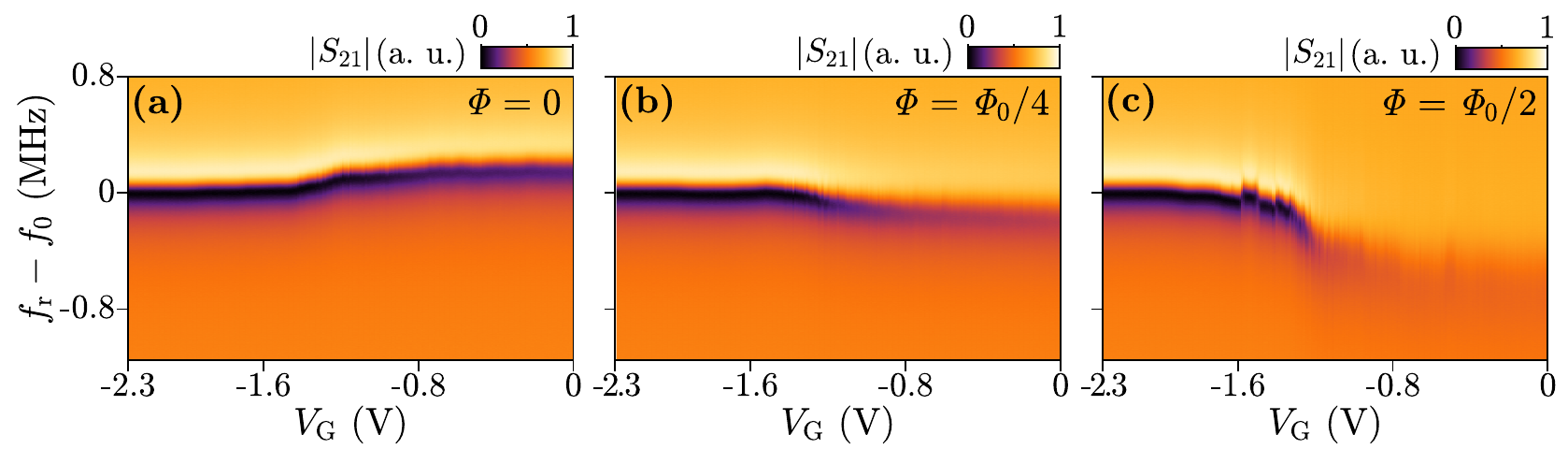}
	\caption{Gate-dependent resonator response at three flux values. Magnitude $\abs{S_{21}}$ of the resonator response was measured around the resonance frequency of the readout resonator while sweeping the gate voltage $V_\mathrm{G}$ at (a) $\it{\Phi}=\mathrm{0}$, (b) $\it{\Phi}=\it{\Phi}_\mathrm{0}/\mathrm{4}$ and (c) $\it{\Phi}=\it{\Phi}_\mathrm{0}/\mathrm{2}$. When the Josephson junction (JJ) was pinched-off ($V_\mathrm{G}=-2.3$~V) the resonance frequency approached $f_\mathrm{0}$, independent of $\it{\Phi}$. Increasing $V_\mathrm{G}$ resulted in a flux-dependent shift $\delta f_\mathrm{0}$ of the bare resonance frequency $f_\mathrm{0}$. The magnitude and sign of $\delta f_\mathrm{0}$ depended on the inductance of the JJ. Moreover, at high $V_\mathrm{G}$ the linewidth of the resonator resonance was larger towards $\it{\Phi}=\it{\Phi}_\mathrm{0}/\mathrm{2}$. In this regime, Andreev bound states existed all over the superconducting gap, thus the resonant resonator signal $P_\mathrm{in}$ was likely able to excite transitions between these subgap states, which therefore increased the linewidth.}
	\label{figS7}
\end{figure}

Sweeping the gate voltage $V_\mathrm{G}$ from $-$2.3~V (JJ pinched-off) to 0~V at three values of flux $\it{\Phi}$ led to different responses of the resonator frequency as depicted in Fig.~\ref{figS7}. For $V_\mathrm{G} \leq \mathrm{-2.3}$~V the JJ was pinched-off, and the resonance frequency approached $f_\mathrm{0}$, independent of $\it{\Phi}$. With increasing $V_\mathrm{G}$ the resonator resonance frequency $f_\mathrm{res.}$ deviated from the bare resonance frequency $f_\mathrm{0}$ by~\cite{Haller}: 
\setcounter{mye}{2}
\begin{equation}
	\delta f_\mathrm{0} \equiv f_\mathrm{res.} - f_\mathrm{0} \approx \frac{8}{\mathrm{\pi}^2} \frac{M^2}{L_\mathrm{R}(L_\mathrm{J}+L_\mathrm{Loop})},
	\label{FrequencyShift}
\end{equation}
where $M \approx 14.7$~pH is the calculated mutual inductance between the resonator and the rf-SQUID using the model of a rectangular loop next to a straight current-carrying wire, $L_\mathrm{R} = L_\mathrm{G}^\mathrm{R} + L_\mathrm{K}^\mathrm{R} \approx 2.33$~nH is the calculated total (geometric and kinetic) inductance of the resonator using conformal mapping techniques~\cite{LK,LG}, $L_\mathrm{J}$ is the inductance of the JJ and $L_\mathrm{Loop} = L_\mathrm{G}^\mathrm{Loop} + L_\mathrm{K}^\mathrm{Loop} \approx 201$~pH is the calculated total inductance of the superconducting loop. The inverse Josephson inductance $L_\mathrm{J}^\mathrm{-1}$ is related to a change in the supercurrent $I_\mathrm{s}$ of the rf-SQUID by: 
\setcounter{mye}{3}
\begin{equation}
	L_\mathrm{J}^\mathrm{-1} = \frac{2\pi}{\it{\Phi}_\mathrm{0}} \frac{\partial I_\mathrm{s}(\varphi)}{\partial \varphi},
	\label{JosephsonInductance}
\end{equation}
where $\varphi=2\pi(\it{\Phi}-L_\mathrm{Loop}I_\mathrm{s}(\varphi))/\it{\Phi}_\mathrm{0}$ is the phase difference across the JJ \cite{JosephsonInductance,PhaseJJ}. The Josephson inductance $L_\mathrm{J}$ can be extracted from the current-phase relation (CPR) [see Fig.~\ref{figS9}(a)]. Therefore, $L_\mathrm{J}$ can be positive or negative, leading to positive or negative frequency shifts $\delta f_\mathrm{0}$ of the bare resonator frequency $f_\mathrm{0}$.

\setcounter{myc}{7}

\setcounter{myc}{8}
\begin{figure}[H]
	\centering
	\includegraphics[width=0.71\columnwidth]{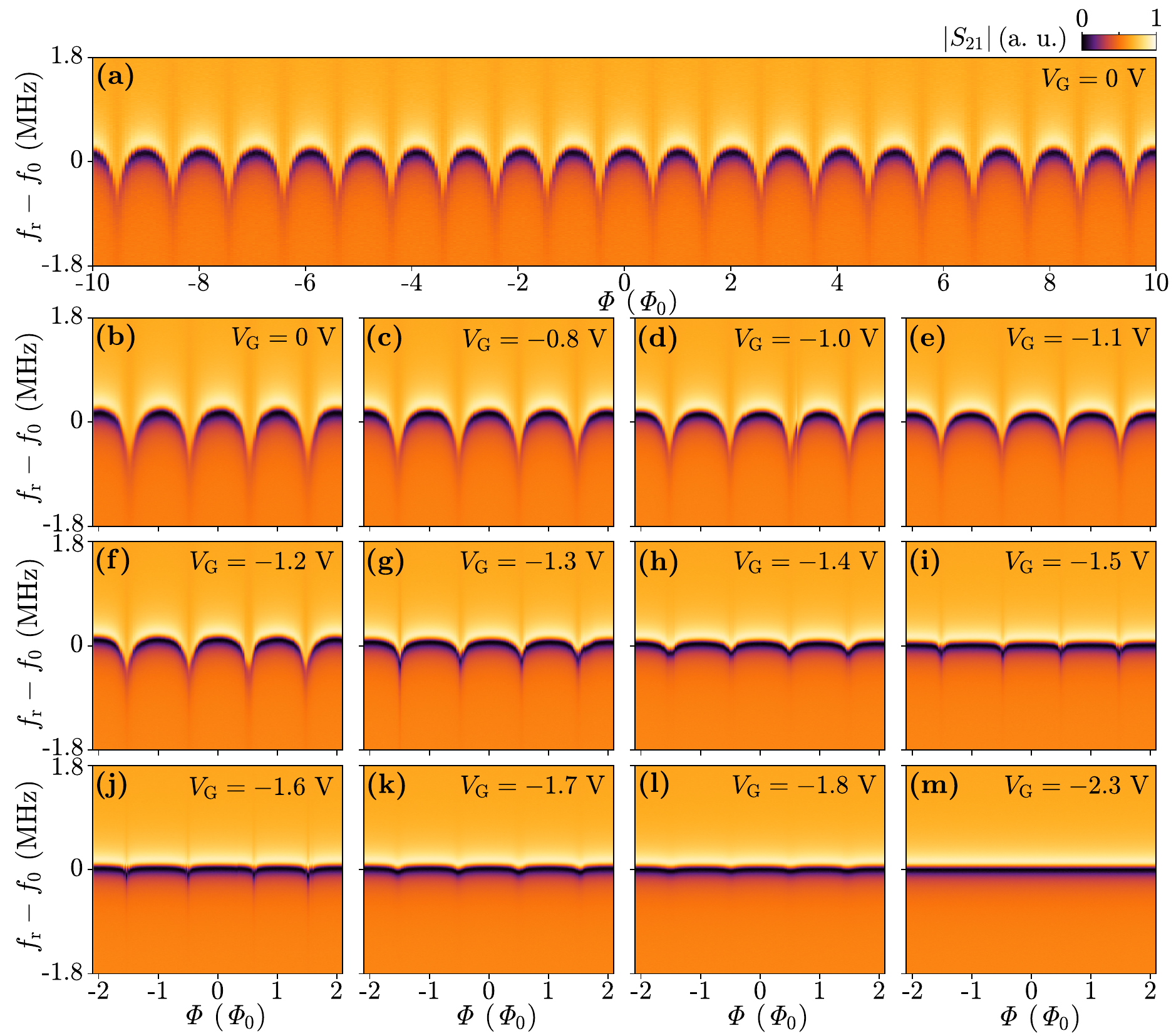}
	\caption{Flux-dependent resonator response at selected gate voltages $V_\mathrm{G}$. (a) Magnitude $\abs{S_{21}}$ was measured around the resonance frequency of the readout resonator as a function of flux $\it{\Phi}$ over 20 periods at $V_\mathrm{G}=\mathrm{0}$~V. (b-m) Flux-dependent magnitude $\abs{S_{21}}$ was monitored in the same frequency range as in (a). The measurements at selected gate voltages $V_\mathrm{G}$ extend the data in Fig.~2(b-e) of the Main Text. These data sets were used to map characteristic quantities of the Josephson junction as a function of $V_\mathrm{G}$, i.e. critical current $I_\mathrm{c}$, effective transmission $\mathrm{\bar{\tau}}$ and Josephson inductance $L_\mathrm{J}$, summarized in Fig.~\ref{figS9}.}
	\label{figS8}
\end{figure}

\setcounter{myc}{9}
\begin{figure}[H]
	\centering
	\includegraphics[width=0.5\columnwidth]{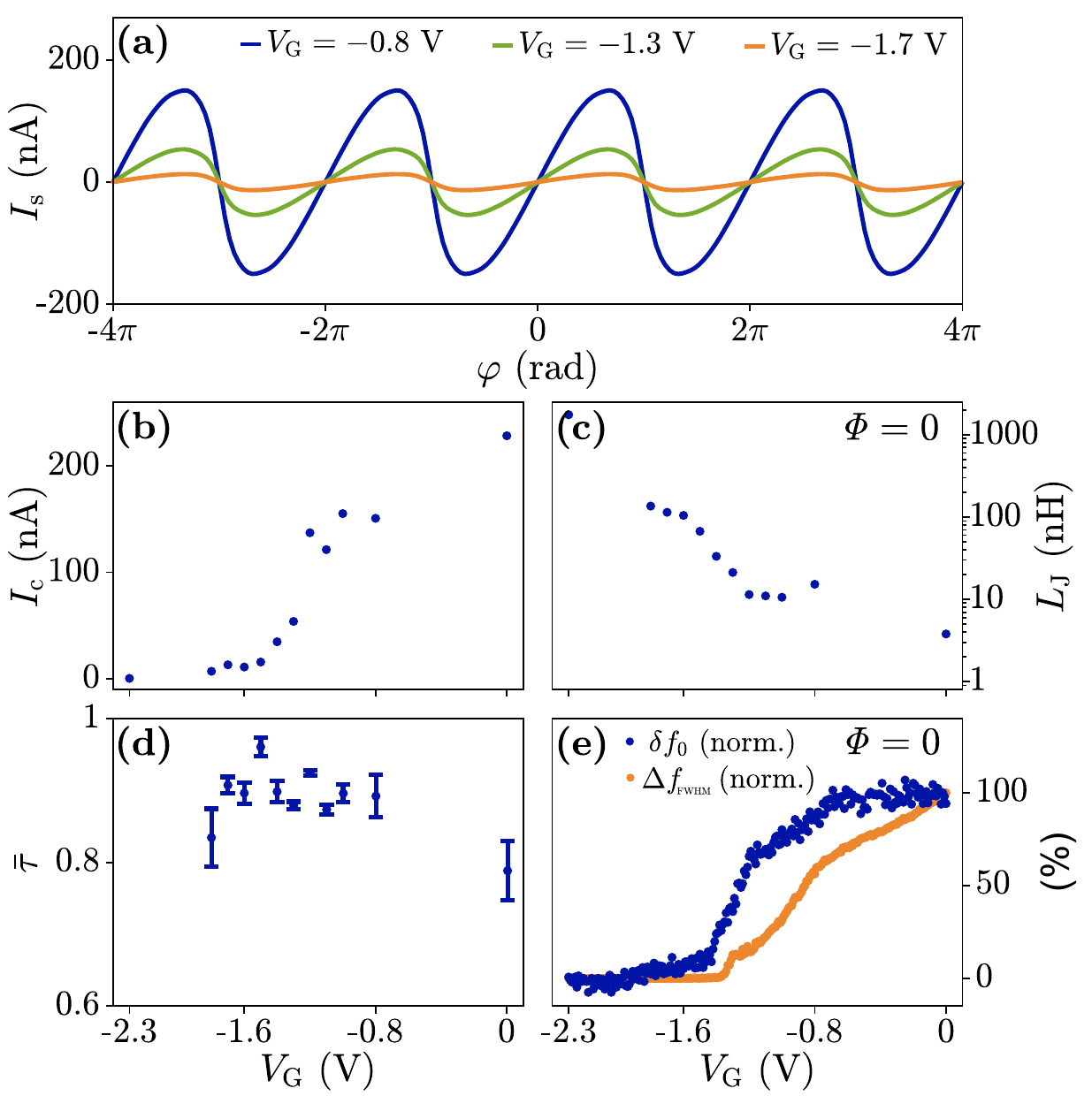}
	\caption{Characteristic quantities of the Josephson junction (JJ) as a function of gate voltages $V_\mathrm{G}$. (a) The current-phase relation (CPR), which is the supercurrent $I_\mathrm{s}$ as a function of phase $\varphi$ across the JJ, is plotted for the three gate voltages $\VG$. The CPR was extracted from flux-dependent single-tone measurements in Fig.~\ref{figS8} using the iterative fitting method described in Ref.~\cite{Haller}. (b) Critical current $I_\mathrm{c}$ as a function of $V_\mathrm{G}$. The $I_\mathrm{c}$ decreased with decreasing $V_\mathrm{G}$, consistent with the reduction of the density of ABSs in the junction. (c) Josephson inductance ($L_\mathrm{J}$) as a function of $V_\mathrm{G}$ at $\it{\Phi}=\mathrm{0}$ was extracted from CPRs using Eq.~\ref{JosephsonInductance}. (d) Effective transmission $\bar{\tau}$, describing the transmission of an effective single ABS replacing all physical ABSs, as a function of $V_\mathrm{G}$ was extracted from CPRs using Eq.~\ref{EffectiveTransmission}. (e) Normalized readout resonator shift $\delta f_\mathrm{0}$ and normalized linewidth of the resonator $\Delta f_\mathrm{FWHM}$ as a function of $V_\mathrm{G}$ at $\it{\Phi}=\mathrm{0}$.}
	\label{figS9}
\end{figure}

By iteratively fitting the periodic shifts of the resonator frequency in Fig.~\ref{figS8} in a self-consistent way under consideration of self-screening effects~\cite{Haller}, the CPR was obtained for each gate voltage $V_\mathrm{G}$. The resulting CPR for three gate voltages, which were used in Fig.~2(b-d) of the Main Text, is shown in Fig.~\ref{figS9}(a). The CPR contains in its amplitude the information about the density of ABSs and in its deviation from sinusoidal shape as an indication of whether high transmission modes are present in the junction~\cite{CPR}. 

The critical current $I_\mathrm{c}$, which is the maximal amplitude of the CPR, as a function of $V_\mathrm{G}$ is depicted in Fig.~\ref{figS9}(b). For gate voltages down to $\VG = -1.2$~V, the junction had a high density of ABSs, supported by the fact that each ABS carries a maximal supercurrent of:
\setcounter{mye}{4}
\begin{equation}
	I_\mathrm{s,i}(\varphi) = -\frac{\mathrm{2e}}{h} \frac{\mathrm{\partial}E_\mathrm{A,i}}{\mathrm{\partial}\varphi}<\mathrm{7~nA},
	\label{EffectiveTransmission}
\end{equation}
where $E_\mathrm{A,i}$ is the energy of the individual ABSs, which depends on its individual transmission~\cite{RelatingABSCPR}. The maximal critical current $I^\mathrm{max}_\mathrm{c}\approx\mathrm{230}$~nA estimated from CPR is comparable to that regularly measured in DC transport experiments~\cite{fornieri2019evidence,haxell2022microwave}. By further decreasing $V_\mathrm{G}$, the critical current was rapidly reduced to $I_\mathrm{c}\approx\mathrm{13}$~nA. Therefore, in this gate voltage regime only few high-transmissive ABSs could be present in the junction. For $\VG < -1.4$~V, $I_\mathrm{c}$ decreased slowly indicating a reduction in the transmission of the individual ABSs rather than its density, which is consistent with observations from two-tone spectroscopy measurements in Fig.~4 of the Main Text. In addition, the Josephson inductance $L_\mathrm{J}$ was extracted from CPR using Eq.~\ref{JosephsonInductance}, which is shown in Fig.~\ref{figS9}(c) for $\it{\Phi}=\mathrm{0}$. 

Since it is not feasible to assign a transparency to each individual ABS, the transmission of the junction was instead described macroscopically with an effective transmission $\mathrm{\bar{\tau}}$, which is identical for all ABSs. The effective transmission $\mathrm{\bar{\tau}}$ was determined by fitting the CPR using:  
\setcounter{mye}{5}
\begin{equation}
	I_\mathrm{s}(\varphi) = I_\mathrm{0}\frac{\mathrm{\bar{\tau}sin(\it{\varphi})}}{\sqrt{\mathrm{1-\bar{\tau}sin^2(\it{\varphi}/\mathrm{2})}}},
	\label{EffectiveTransmission}
\end{equation}
where $I_\mathrm{0}=(\mathrm{e}/\mathrm{2}\hbar)\bar{N}\Delta$, with $\bar{N}$ is the effective number of ABSs in the junction~\cite{EffectiveTransmissionFormula}. Figure~\ref{figS9}(d) summarizes the $\mathrm{\bar{\tau}}$ as a function of $V_\mathrm{G}$. Across the full gate voltage range a high $\mathrm{\bar{\tau}}$ was observed, which is consistent with the interpretation of the junction being in the diffusive long regime for $\VG > -1.4$~V and ballistic short regime for $\VG < -1.4$~V, respectively. In the diffusive regime, there are many ABSs with low transmission as well as high transmission following Dorokhov's bimodal distribution~\cite{Dorokhov,TransmissionDistribution}, resulting in a relatively high $\mathrm{\bar{\tau}}$. For $\VG < -1.4$~V, the junction hosted only few ABSs as discussed in Fig.~\ref{figS9}(b). Two-tone spectroscopy measurements in Fig.~4 of the Main Text revealed that some of the ABSs had high transmission, but there were also some with low transmission, so that $\mathrm{\bar{\tau}}$ remained approximately constant. 

The resonance frequency shift as a function of $\VG$, together with $\Qi$, in Fig.~2(a) of the Main Text revealed more details about different gate voltage regimes. Figure~\ref{figS9}(e) shows the percentage change of both the frequency shift $\delta f_\mathrm{0}$ and the linewidth $\Delta f_\mathrm{FWHM}$ of the resonator resonance as a function of $\VG$ at $\it{\Phi}=\mathrm{0}$. For gate voltages down to $\VG = -0.8$~V, the resonance frequency remained nearly unaffected because ABSs with low transmission were depleted, which had a minor contribution to the supercurrent. However, they likely acted as loss channels and thus $\Delta f_\mathrm{FWHM}$ decreased. As $\VG$ was tuned to more negative values, both $\delta f_\mathrm{0}$ and $\Delta f_\mathrm{FWHM}$ decreased, indicating a further reduction of the ABS density in the junction. For $\VG < -1.4$~V, the resonance frequency still decreased while the linewidth of the resonator remained unchanged. This is attributed to the absence of available ABSs for microwave-induced transitions at $\it{\Phi}=\mathrm{0}$ and a reduction in the transmission of individual ABSs. This interpretation is consistent with the two-tone spectroscopy data in Figs.~3 and~4 of the Main Text.

\setcounter{myc}{10}
\begin{figure}[H]
	\centering
	\includegraphics[width=0.5\columnwidth]{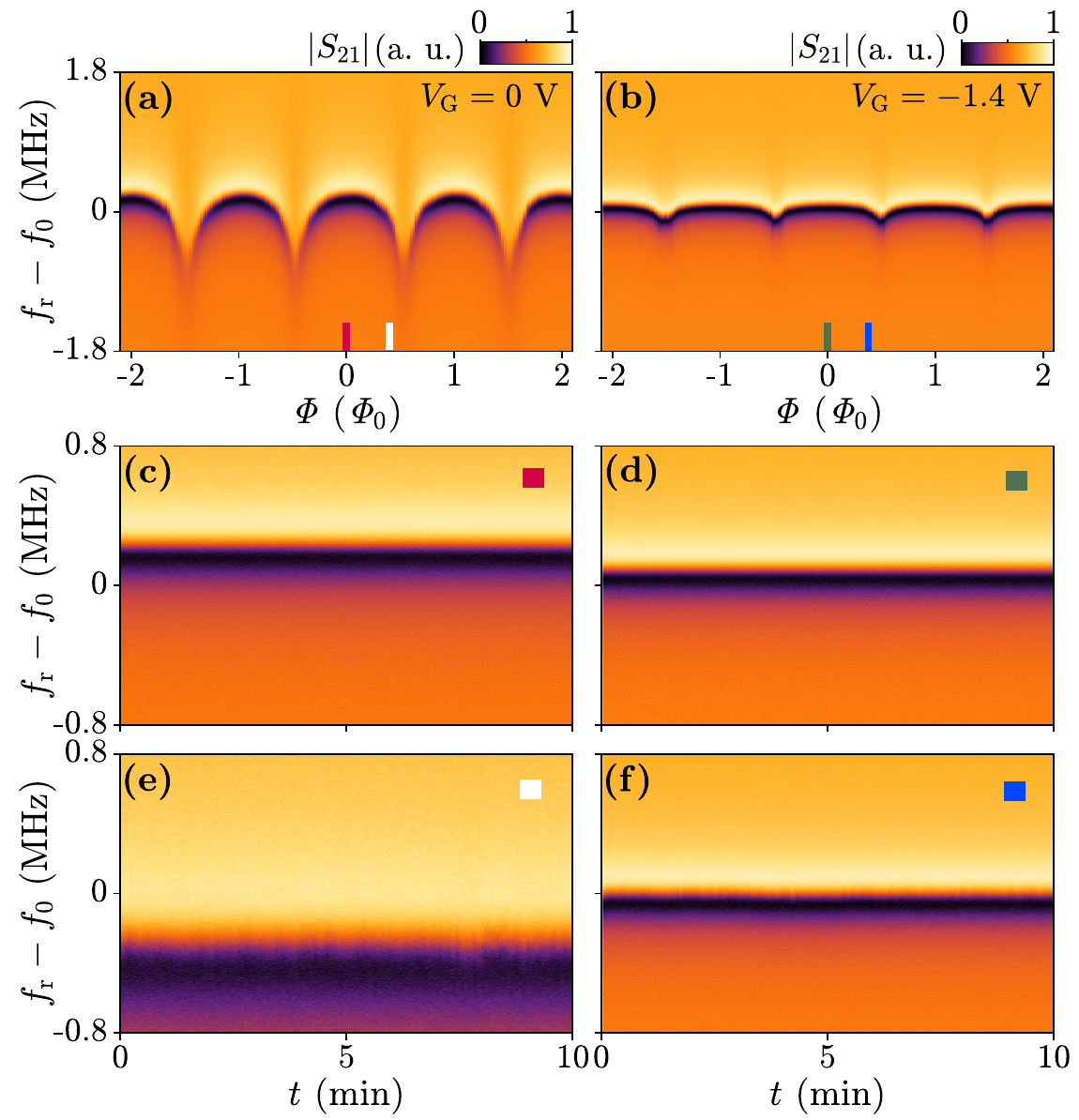}
	\caption{Time stability of the resonator response at gate voltage $V_\mathrm{G}$ and flux $\it{\Phi}$ sensitive points. Magnitude $\abs{S_{21}}$ was monitored around the resonance frequency of the readout resonator as a function of flux $\it{\Phi}$ at (a) $V_\mathrm{G}=\mathrm{0}$~V and (b) $V_\mathrm{G}=\mathrm{-1.4}$~V to illustrate the configuration at which the data in (c-f) were taken, indicated by the colored bars. Evolution of the magnitude $\abs{S_{21}}$ around the resonance frequency for (c) $V_\mathrm{G}=\mathrm{0}$~V and (d) $V_\mathrm{G}=\mathrm{-1.4}$~V within 10 minutes. (e, f) Same as (c, d) at a flux $\it{\Phi}$, at which the resonance frequency shifted remarkably for varying $\it{\Phi}$. Absence of fluctuations and drifts in the resonance frequency at the flux insensitive point $\it{\Phi}=\mathrm{0}$ for both $V_\mathrm{G}=\mathrm{0}$~V and $V_\mathrm{G}=\mathrm{-1.4}$~V suggested gate voltage induced charge noise was negligible. The resonance frequency stayed stable for periods over 10 minutes at the flux sensitive point in (e) and (f), indicating that there were no magnetic field drifts observable during that time. Small fluctuations in the resonator response on a shorter time scale caused by magnetic field fluctuations were observed. However, they did not limit the two-tone spectroscopy measurements in this work, since these were performed around the flux insensitive point $\it{\Phi}=\it{\Phi}_\mathrm{0}/\mathrm{2}$.}
	\label{figS10}
\end{figure}

\subsection{Additional two-tone spectroscopy measurements for two selected drive powers}
\setcounter{myc}{11}
\begin{figure}[H]
	\centering
	\includegraphics[width=0.5\columnwidth]{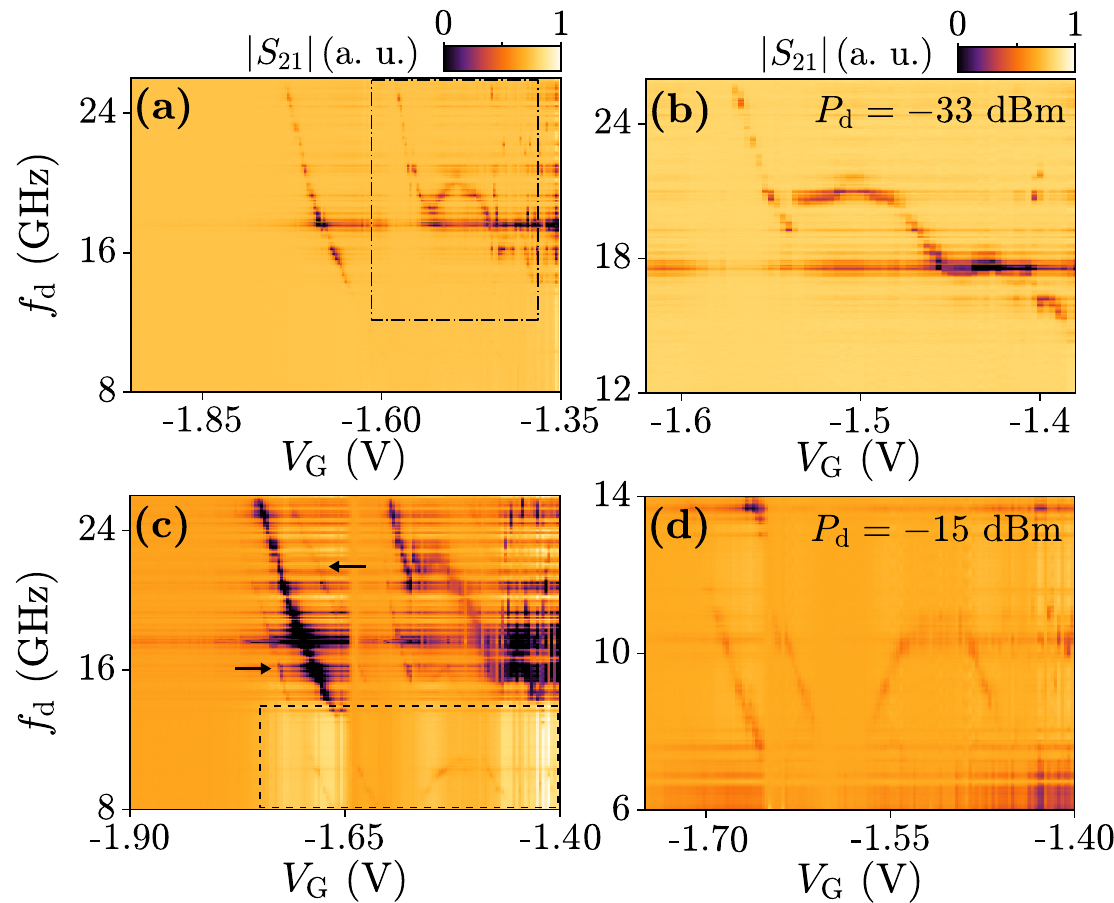}
	\caption{Gate-dependent two-tone spectroscopy measurements for two selected drive powers. (a) Magnitude $\abs{S_{21}}$ of the readout resonator response as a function of drive tone frequency $\fd$ at $\it{\Phi} = \it{\Phi}_\mathrm{0}/\mathrm{2}$, as well as gate voltage $\VG$. The drive tone power was set to $P_\mathrm{d}=\mathrm{-33}$~dBm. (b) Probing the region marked with a box in (a) with higher resolution. (c) Same as (a) with $P_\mathrm{d}=\mathrm{-15}$~dBm. (d) Zoomed-in measurement of the region marked with a box in (c). Horizontal lines in the two-tone spectroscopy measurements were present predominantly above $\fd=14$~GHz and only at gate voltage and flux values at which an Andreev bound state (ABS) pair transition was observed (see also Fig.~3 of the Main Text). Moreover, within this range the strength of the horizontal lines remained constant as a function of $\VG$ and $\it{\Phi}$, but increased at stronger driving powers $P_\mathrm{d}$. We attribute these horizontal lines to resonances in the drive line and connected measurement circuit~\cite{NanowireLongJunction6}. Stronger drive powers and the unintentional resonances in the drive line were responsible for the increased linewidth of the ABS pair transition spectrum~\cite{PowerBroadening}. Replicas of ABSs, marked with arrows in panel~(c), were observed at strong driving powers. Following Ref.~\cite{FlipChip4}, we assign these replicas to transitions involving absorption or emission of a resonator photon.}
	\label{figS11}
\end{figure}

\subsection{Additional data for the study of electrostatic tuning of isolated Andreev bound states}

\setcounter{myc}{12}
\begin{figure}[H]
	\centering
	\includegraphics[width=0.5\columnwidth]{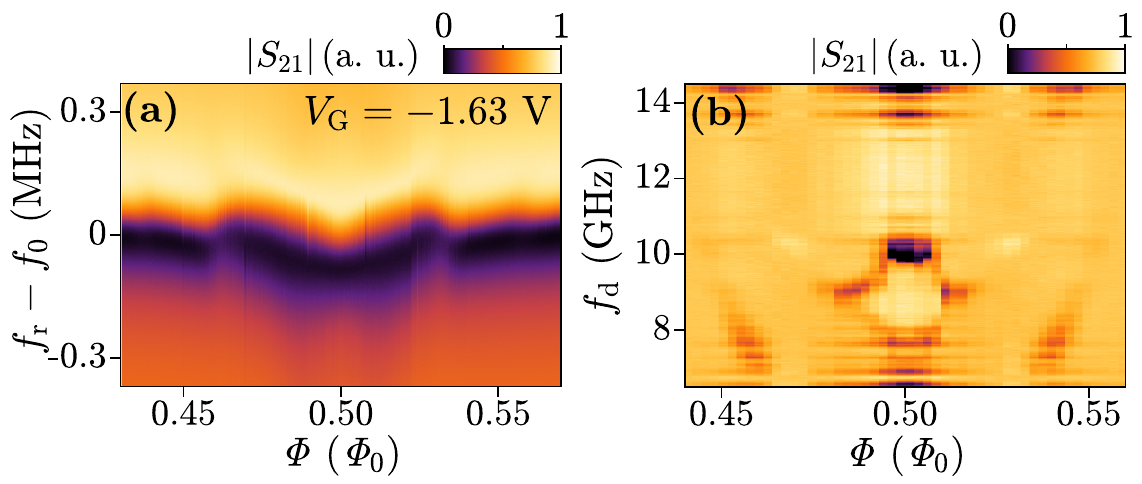}
	\caption{Flux-dependent single- and two-tone spectroscopy data in the vicinity of the incoherent resonance at $\VG=\mathrm{-1.63}$~V as discussed in Fig.~4 of the Main Text. (a) Frequency response of the readout resonator as a function of flux $\it{\Phi}$ at $\VG=\mathrm{-1.63}$~V. An avoided crossing, symmetric around $\it{\Phi}=\it{\Phi}_\mathrm{0}/\mathrm{2}$, was observed. Following Ref.~\cite{Resonator2DEG2}, this avoided crossing is a sign of virtual resonator-Andreev bound state (ABS) photon exchange, described by a Jaynes-Cummings interaction. (b) Two-tone spectroscopy data corresponding to the single-tone data in (a). The ABS pair transition at $\it{\Phi}=\it{\Phi}_\mathrm{0}/\mathrm{2}$ with $\fd=10$~GHz is consistent with observations in Fig.~4 of the Main Text and Fig.~\ref{figS11}. In the vicinity of $\it{\Phi}=\it{\Phi}_\mathrm{0}/\mathrm{2}$, the transition frequency was reduced due to the incoherent resonance before it increased again.}
	\label{figS12}
\end{figure}

\setcounter{myc}{13}
\begin{figure}[H]
	\centering
	\includegraphics[width=0.56\columnwidth]{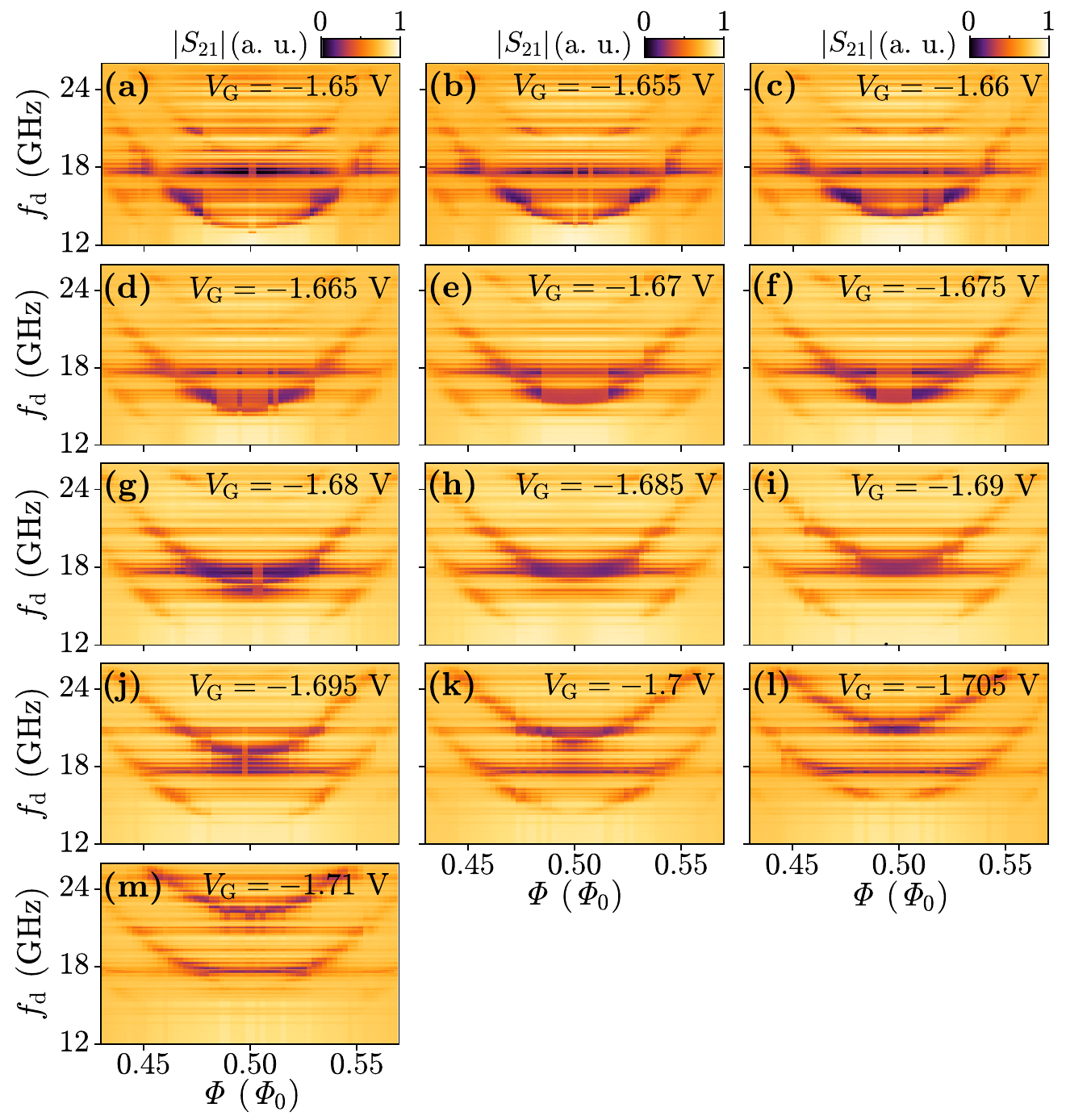}
	\caption{Flux-dependent two-tone spectroscopy measurements at selected gate voltages $\VG$. (a-m) Magnitude $\abs{S_{21}}$ of the readout resonator response as a function of drive tone frequency $\fd$ and flux $\it{\Phi}$. These data sets were used to extract the induced superconducting gaps $\Delta$ and transmissions $\tau$, which are plotted in Fig.~4(d) of the Main Text. By comparing the data with the results in Fig.~\ref{figS11}, replicas of the ABS pair transition were identified and removed from the data set.}
	\label{figS13}
\end{figure}

\section{Estimate of characteristic time scales for qubits}
We performed time-stability measurements at the gate voltage and flux sensitive points, as shown in Fig.~\ref{figS10}. Results indicate the absence of drift and large fluctuations over time scales of several minutes.

We estimated both relaxation time $T_1$ and dephasing time $T_2$ of an Andreev level qubit from our two-tone spectroscopy data. Relaxation of the qubit can be caused by several factors, for example dielectric losses, charge noise, flux noise, quasiparticle fluctuations and Purcell noise. Therefore, all contributions add to $T_1$ as: 
\setcounter{mye}{6}
\begin{equation}
	\frac{1}{T_1}=\frac{1}{T^D}+\frac{1}{T^{N_g}}+\frac{1}{T^{\it{\Phi}}}+\frac{1}{T^{QP}}+\frac{1}{T^P}. 
\end{equation}
Assuming that the main source of relaxation at zero magnetic field is due to dielectric losses~\cite{DielectricLoss1,DielectricLoss2}, the relaxation time can be estimated as given by Eq.~\ref{eqT1}:

\setcounter{mye}{7}
\begin{equation}
	T_1 \approx T^D=\frac{\Qi}{2\pi f_Q},
	\label{eqT1}
\end{equation}
where $\Qi$ is the internal quality factor of the resonator. We find $T_1 \approx 600$~ns for $\Qi=5\cdot10^4$ and $f_Q=14$~GHz, corresponding to the resonator used for the measurements and the minimum frequency of the ABS pair transition shown in Fig.~3(c). Possible attempts to improve the lifetime $T_1$ could include using materials with lower loss tangent, providing better shielding in the device environment and optimizing the fabrication of the Resonator Chip.

\setcounter{myc}{14}
\begin{figure}[H]
	\centering
	\includegraphics[width=0.25\columnwidth]{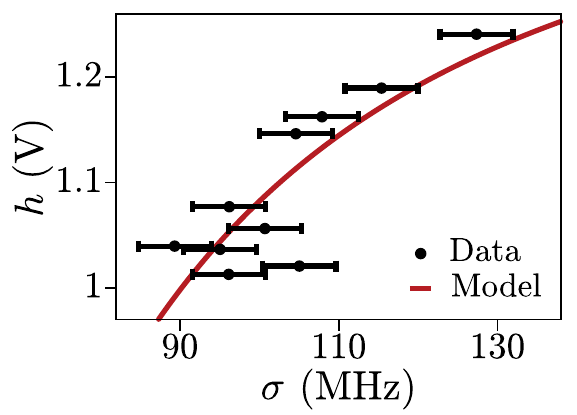}
	\caption{Parametric plot of spectral line amplitude $h$ from two-tone spectroscopy as a function of its linewidth $\sigma$ (drive power) together with a fit to theory prediction. Zero-drive power linewidth $\sigma_\mathrm{min}$ can be estimated as a point where theory line intercepts $h$=0.}
	\label{figS14}
\end{figure}

The inhomogeneous dephasing time $T_2^*$ is given by: 
\setcounter{mye}{8}
\begin{equation}
	\frac{1}{T_2^*}=\frac{1}{2T_1}+\frac{1}{T_{\phi}},
\end{equation}

where $T_{\phi}$ is the pure dephasing time which characterizes qubit frequency fluctuations. For the estimation of the inhomogeneous dephasing time $T_2^*$ we followed the technique outlined in Ref.~\cite{Petersson_linewidth}, where spectral line amplitude $h$ and linewidth $\sigma$ were tracked as a function of drive power $P_d$ and fitted with $h=h_\mathrm{0}(1-\sigma_\mathrm{min}^2/\sigma^2)$ as shown in Fig.~\ref{figS14}. The fit can be used to estimate $T_2^*$ using zero drive power parameter $\sigma_\mathrm{min}$ as  $T_2^*=1/(\sqrt{2}\pi \sigma_\mathrm{min}) \approx$~4.5~ns. Pure dephasing could be reduced by better filtering of input signals and shielding of the sample space from the environment. 

\end{document}